\documentclass[a4paper,12pt,twoside]{article}
\pdfoutput=1
\usepackage{abstract}

\usepackage[utf8]{inputenc}
\usepackage[T1]{fontenc}

\usepackage[english]{babel}

\usepackage{amsmath}
\usepackage{graphicx}
\usepackage{hyperref}
\hypersetup{colorlinks=true, allcolors=blue,pdfproducer={}}
\usepackage{fancyhdr}
\usepackage{wrapfig}
\usepackage{subcaption}
\usepackage[braket]{qcircuit}
\usepackage{url}
\usepackage{siunitx}
\usepackage{multirow}

\usepackage{placeins}

\usepackage{ifthen}
\newboolean{uprightparticles}
\setboolean{uprightparticles}{false}
\usepackage{xspace} 
\usepackage{upgreek}

\def\lhcb   {\mbox{LHCb}\xspace}

\def\MagUp {\mbox{\em Mag\kern -0.05em Up}\xspace}

\ifthenelse{\boolean{uprightparticles}}%
{

 \def\Ppi         {\ensuremath{\uppi}\xspace}

 \def\PDelta      {\ensuremath{\Delta}\xspace}                 
 \def\PXi         {\ensuremath{\Xi}\xspace}                 
 \def\PLambda     {\ensuremath{\Lambda}\xspace}                 
 \def\PSigma      {\ensuremath{\Sigma}\xspace}                 
 \def\POmega      {\ensuremath{\Omega}\xspace}                 
 \def\PUpsilon    {\ensuremath{\Upsilon}\xspace}

 \def\PB      {\ensuremath{\mathrm{B}}\xspace}                 
                  
 \def\PD      {\ensuremath{\mathrm{D}}\xspace}

 \def\PK      {\ensuremath{\mathrm{K}}\xspace}

 \def\Pb      {\ensuremath{\mathrm{b}}\xspace}                 
 \def\Pc      {\ensuremath{\mathrm{c}}\xspace}

 \def\Pi      {\ensuremath{\mathrm{i}}\xspace}

 \def\Pp      {\ensuremath{\mathrm{p}}\xspace}

 \def\Ps      {\ensuremath{\mathrm{s}}\xspace}                 
                  
 \def\Pu      {\ensuremath{\mathrm{u}}\xspace}

 \def\thebaroffset{0.0em}
}
{

 \def\Ppi         {\ensuremath{\pi}\xspace}

 \mathchardef\PDelta="7101
 \mathchardef\PXi="7104
 \mathchardef\PLambda="7103
 \mathchardef\PSigma="7106
 \mathchardef\POmega="710A
 \mathchardef\PUpsilon="7107
                  
 \def\PB      {\ensuremath{B}\xspace}                 
                  
 \def\PD      {\ensuremath{D}\xspace}

 \def\PK      {\ensuremath{K}\xspace}

 \def\Pb      {\ensuremath{b}\xspace}                 
 \def\Pc      {\ensuremath{c}\xspace}

 \def\Pi      {\ensuremath{i}\xspace}

 \def\Pp      {\ensuremath{p}\xspace}

 \def\Ps      {\ensuremath{s}\xspace}                 
                  
 \def\Pu      {\ensuremath{u}\xspace}

 \def\thebaroffset{0.18em}
}
\newcommand{\offsetoverline}[2][\thebaroffset]{\kern #1\overline{\kern -#1 #2}}%

\DeclareRobustCommand{\optbar}[1]{\shortstack{{\miniscule (\rule[.5ex]{1.25em}{.18mm})}
  \\ [-.7ex] $#1$}}

\def\uquark    {{\ensuremath{\Pu}}\xspace}

\def\squark    {{\ensuremath{\Ps}}\xspace}

\def\cquark    {{\ensuremath{\Pc}}\xspace}

\def\bquark    {{\ensuremath{\Pb}}\xspace}

\def\pion   {{\ensuremath{\Ppi}}\xspace}

\def\pip    {{\ensuremath{\pion^+}}\xspace}
\def\pim    {{\ensuremath{\pion^-}}\xspace}

\def\kaon    {{\ensuremath{\PK}}\xspace}

\def\KorKbar {\kern \thebaroffset\optbar{\kern -\thebaroffset \PK}{}\xspace}

\def\Kp      {{\ensuremath{\kaon^+}}\xspace}
\def\Km      {{\ensuremath{\kaon^-}}\xspace}

\def\D       {{\ensuremath{\PD}}\xspace}

\def\DorDbar {\kern \thebaroffset\optbar{\kern -\thebaroffset \PD}\xspace}

\def\Dp      {{\ensuremath{\D^+}}\xspace}
\def\Dm      {{\ensuremath{\D^-}}\xspace}

\def\DpDm    {\ensuremath{\Dp {\kern -0.16em \Dm}}\xspace}

\def\Dsp     {{\ensuremath{\D^+_\squark}}\xspace}
\def\Dsm     {{\ensuremath{\D^-_\squark}}\xspace}

\def\B       {{\ensuremath{\PB}}\xspace}

\def\BorBbar {\kern \thebaroffset\optbar{\kern -\thebaroffset \PB}\xspace}
\def\Bz      {{\ensuremath{\B^0}}\xspace}

\def\Bd      {{\ensuremath{\B^0}}\xspace}

\def\BdorBdbar {\kern \thebaroffset\optbar{\kern -\thebaroffset \Bd}\xspace}

\def\Bs      {{\ensuremath{\B^0_\squark}}\xspace}

\def\BsorBsbar {\kern \thebaroffset\optbar{\kern -\thebaroffset \Bs}\xspace}

\def\Y#1S{\ensuremath{\PUpsilon{(#1S)}}\xspace}

\def\proton      {{\ensuremath{\Pp}}\xspace}

\def\Lz          {{\ensuremath{\PLambda}}\xspace}

\def\LorLbar     {\kern \thebaroffset\optbar{\kern -\thebaroffset \PLambda}\xspace}

\def\Lc          {{\ensuremath{\Lz^+_\cquark}}\xspace}

\def\Lb           {{\ensuremath{\Lz^0_\bquark}}\xspace}

\def\BF         {{\ensuremath{\mathcal{B}}}\xspace}

\newcommand{\decay}[2]{\ensuremath{#1\!\to #2}\xspace} 

\def\to                 {\ensuremath{\rightarrow}\xspace}

\def\CP                {{\ensuremath{C\!P}}\xspace}

\def\Vub  {{\ensuremath{V_{\uquark\bquark}^{\phantom{\ast}}}}\xspace}

\def\AT#1     {\ensuremath{A_{\mathrm{T}}^{#1}}\xspace}           

\def\C#1      {\ensuremath{\mathcal{C}_{#1}}\xspace}                       
\def\Cp#1     {\ensuremath{\mathcal{C}_{#1}^{'}}\xspace}                    
\def\Ceff#1   {\ensuremath{\mathcal{C}_{#1}^{\mathrm{(eff)}}}\xspace}        
\def\Cpeff#1  {\ensuremath{\mathcal{C}_{#1}^{'\mathrm{(eff)}}}\xspace}       
\def\Ope#1    {\ensuremath{\mathcal{O}_{#1}}\xspace}                       
\def\Opep#1   {\ensuremath{\mathcal{O}_{#1}^{'}}\xspace}                    

\newcommand{\aunit}[1]{\ensuremath{\text{\,#1}}}       

\newcommand{\tev}{\aunit{Te\kern -0.1em V}\xspace}
\newcommand{\gev}{\aunit{Ge\kern -0.1em V}\xspace}
\newcommand{\mev}{\aunit{Me\kern -0.1em V}\xspace}
\newcommand{\kev}{\aunit{ke\kern -0.1em V}\xspace}
\newcommand{\ev}{\aunit{e\kern -0.1em V}\xspace}
 
\newcommand{\mevc}{\ensuremath{\aunit{Me\kern -0.1em V\!/}c}\xspace}
\newcommand{\gevc}{\ensuremath{\aunit{Ge\kern -0.1em V\!/}c}\xspace}
\newcommand{\mevcc}{\ensuremath{\aunit{Me\kern -0.1em V\!/}c^2}\xspace}
\newcommand{\gevcc}{\ensuremath{\aunit{Ge\kern -0.1em V\!/}c^2}\xspace}

\def\fb   {\ensuremath{\aunit{fb}}\xspace}
\def\invfb   {\ensuremath{\fb^{-1}}\xspace}

\def\gsim{{~\raise.15em\hbox{$>$}\kern-.85em
          \lower.35em\hbox{$\sim$}~}\xspace}
\def\lsim{{~\raise.15em\hbox{$<$}\kern-.85em
          \lower.35em\hbox{$\sim$}~}\xspace}

\def\sqs   {\ensuremath{\protect\sqrt{s}}\xspace}

\def\evtgen     {\mbox{\textsc{EvtGen}}\xspace}

\def\geant      {\mbox{\textsc{Geant4}}\xspace}

\def\pythia     {\mbox{\textsc{Pythia}}\xspace}

\def\tell1  {TELL1\xspace}
\def\ukl1   {UKL1\xspace}

\newcommand{\BdDsPi}   {\texorpdfstring{\decay{\Bz}{\Dsp \pim}}{}}

\newcommand{\DsKKPi}   {\decay{\Dsm}{\Km\Kp\pim}}

\newcommand{\LbDsP}    {\texorpdfstring{\decay{\Lb}{\Dsm p}}{}}

\newcommand{\LbLcPi}   {\decay{\Lb}{\Lc \pim}}

\newcommand{\LcPKPi}   {\decay{\Lc}{\proton \Km \pip}}

\newcommand{\LcPi}     {\texorpdfstring{\Lc \pim}{}}

\newcommand{\BFLbDspPaper}{(12.6 \pm 0.5 \pm 0.3 \pm 1.2 )\times 10^{-6}}

\newcommand{\RatioBFPAPER}{(2.56 \pm 0.10 \pm 0.05 \pm 0.14 )\times 10^{-3}}

\newcommand{\YieldLbDsp}{831\pm32}

\newcommand{\YieldLbLcPiPaper}{(4.047\pm0.007)\times 10^5}

\newcommand{\EffLbDspPerc}{0.1819 \pm 0.0013}

\newcommand{\EffLbLcPiPerc}{0.1947 \pm 0.0012}

\usepackage{amsmath}
\usepackage{subcaption}
\usepackage{booktabs}

\title{Proceedings of the IFJ PAN\\Particle Physics Summer Student\\Alumni Conference 2023\\(Kraków, 14 -- 15 July 2023)}
\author{Dominik Derendarz, Rafał Staszewski, Maciej Trzebiński}
\date{December 2023}

\input{papers.cls}

\newcommand{\ResCnt}{\setcounter{figure}{0} \setcounter{table}{0} \setcounter{section}{0} \setcounter{equation}{0}}

\makeatletter
\def\@maketitle{%
    \newpage
    \null
    \vskip 1em%
    \begin{center}%
    \includegraphics[width=1.0\textwidth]{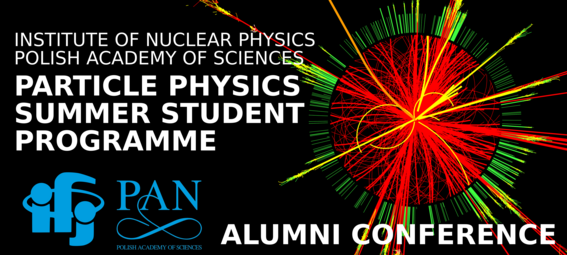}
    \let \footnote \thanks
    \vskip 3em%
    {\linespread{1.2}\Huge\bfseries \@title \par}%
    \vskip 6em%
    {\Large Edited by:\\[2pt]
      \begin{tabular}[t]{c}%
        \@author
      \end{tabular}\par}%
    \vskip 7em%
    {\Large \@date}
  \end{center}
  \par
  \vskip 1.5em}
\makeatother

\begin{document}

\begin{titlepage}
  \maketitle
  \thispagestyle{empty}
  \setcounter{page}{0}
\end{titlepage}

\newpage
\thispagestyle{empty}
\vspace*{15cm}
\noindent Published by The Henryk Niewodnicza\'nski Institute of Nuclear Physics Polish Academy of Sciences\\[0.4cm]
Krak\'ow 2023\\[0.4cm]
Reviewers: dr Domink Derendarz, dr Rafa{\l} Staszewski, dr Maciej Trzebi\'nski\\[0.4cm]
ISBN: 978-83-63542-39-9\\
DOI: \href{https://doi.org/10.48733/978-83-63542-39-9}{https://doi.org/10.48733/978-83-63542-39-9}
\newpage

\setcounter{tocdepth}{0}
\thispagestyle{empty}
\tableofcontents 
\clearpage

\newpage
\thispagestyle{empty}
\mbox{\ }
\setcounter{page}{1}
\newpage

\section*{Foreword}

The second edition Particle Physics Summer Student Alumni Conference was held in July, 14-15 2023 at the Henryk Niewodniczański Institute of Nuclear Physics Polish Academy of Science. The history of PPSS programme is described in the proceedings from 1$^{st}$ PPSS Alumni Conference \cite{PPSS_proc}. In this edition 18 alumni gave talks. There was also a ``motivational talk'' given by dr Maciej Trzebi\'nski and advertisement of Krakow Interdisciplinary Doctoral School by Ms. Aleksandra Pacanowska. A very important point in the programme were discussion sessions devoted to Msc. and PhD. studies and scientific career. The following talks were given:
\begin{itemize}
    \item Maciej Giza, \textit{First observation and branching fraction measurement of the $\Lambda^0_b \to D_s^-p$ decay},
    \item Miguel Ruiz Diaz, \textit{Towards a new test of lepton flavor universality using $B_0 \to K^*_0e^+e^-$ decays in the high di-lepton invariant mass region},
    \item Ferhat Ozturk, \textit{Alignment of SiT detector in ATLAS Forward Proton detectors},
    \item Viktoriia Lysenko, \textit{Studies of ATLAS Forward Proton (AFP) ToF performance with Run-3 data},
    \item Patrycja Potępa, \textit{Hunting for jets in proton-lead collisions in the ATLAS experiment at the LHC},
    \item Szymon Sławiński, \textit{The GRAND Quest for Neutrino Astronomy with Giant Radio Array for Neutrino Detection},
    \item Marouane Benhassi, \textit{Measuring the quantum efficiency of photomultipliers at the CAPACITY laboratory for the KM3NeT experiment},
    \item Sara Ruiz Daza, \textit{Monte Carlo Simulations of Detector Prototypes Designed in a 65 nm CMOS Imaging Process},
    \item Paula Erland, \textit{Evidence of the Exclusive Jet Production Using the ATLAS Detector},
    \item Pragati Patel, \textit{Diffractive Physics and Beam Optics at LHC},
    \item Clarisse Prat, \textit{Analysis of single-diffractive production of charmed mesons in ATLAS},
    \item Aryan Borkar, \textit{The measurement of $\phi(1020)$ meson production in proton-proton collisions at $\sqrt{s} = 13.6$ TeV with the ATLAS detector at the LHC},
    \item Alexia Mavrantoni, \textit{Optical trapping using evanescent field},
    \item Mateusz Kmieć, \textit{Formalism in CPT Studies with (Charm) Neutral Mesons and the Standard Model Extension},
    \item Mateusz Zych, \textit{The Bubbling Universe: Cosmological Phase Transitions and Their Consequences},
    \item Giorgi Asatiani, \textit{Cosmic-Ray Cascades Development - Asymmetries and Fluctuations},
    \item Karolina Kmieć, \textit{Dark Matter in Confining Dark Sectors},
    \item Sebastian Dawid, \textit{Three-particle scattering amplitudes from Lattice QCD}.
\end{itemize}

Except alumni, in conference participated students from 11$^{th}$ edition of PPSS programme.

\subsection*{PPSS 2023}
A record number of 181 candidatures were submitted, from which 32 students were accepted. Finally, we had participants from 12 countries\footnote{Poland, Colombia, Greece, Hungary, Italy, Lithuania, Peru, Serbia, Spain, Sweden, Turkey and United Kingdom.}. Thanks to the support from IFJ PAN and Polish Academy of Sciences we were able to fully cover the cost of stay at dorms for 27 participants!

\begin{figure}[!htbp]
    \centering
    \includegraphics[width=0.32\textwidth]{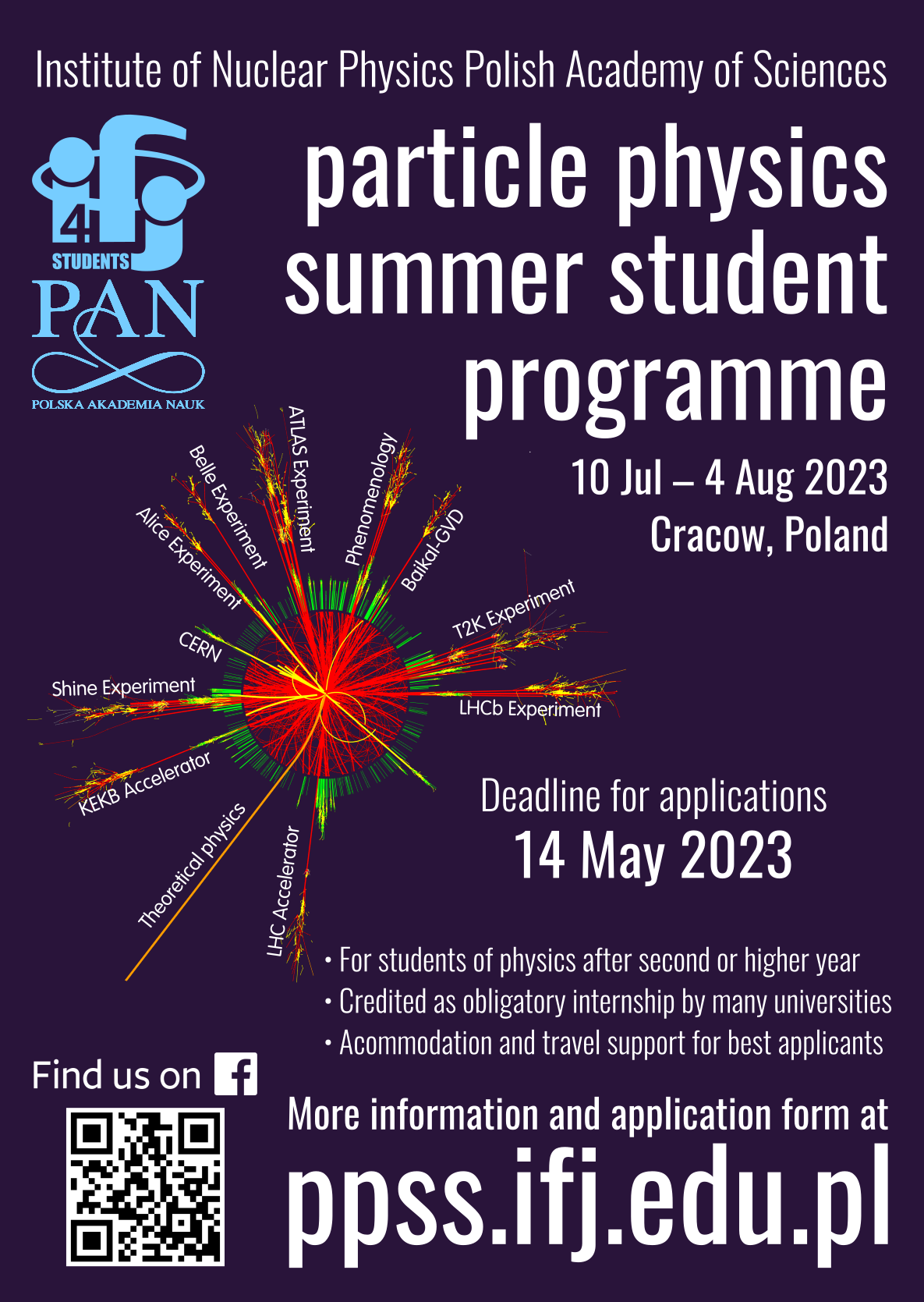}
    \includegraphics[width=0.67\textwidth]{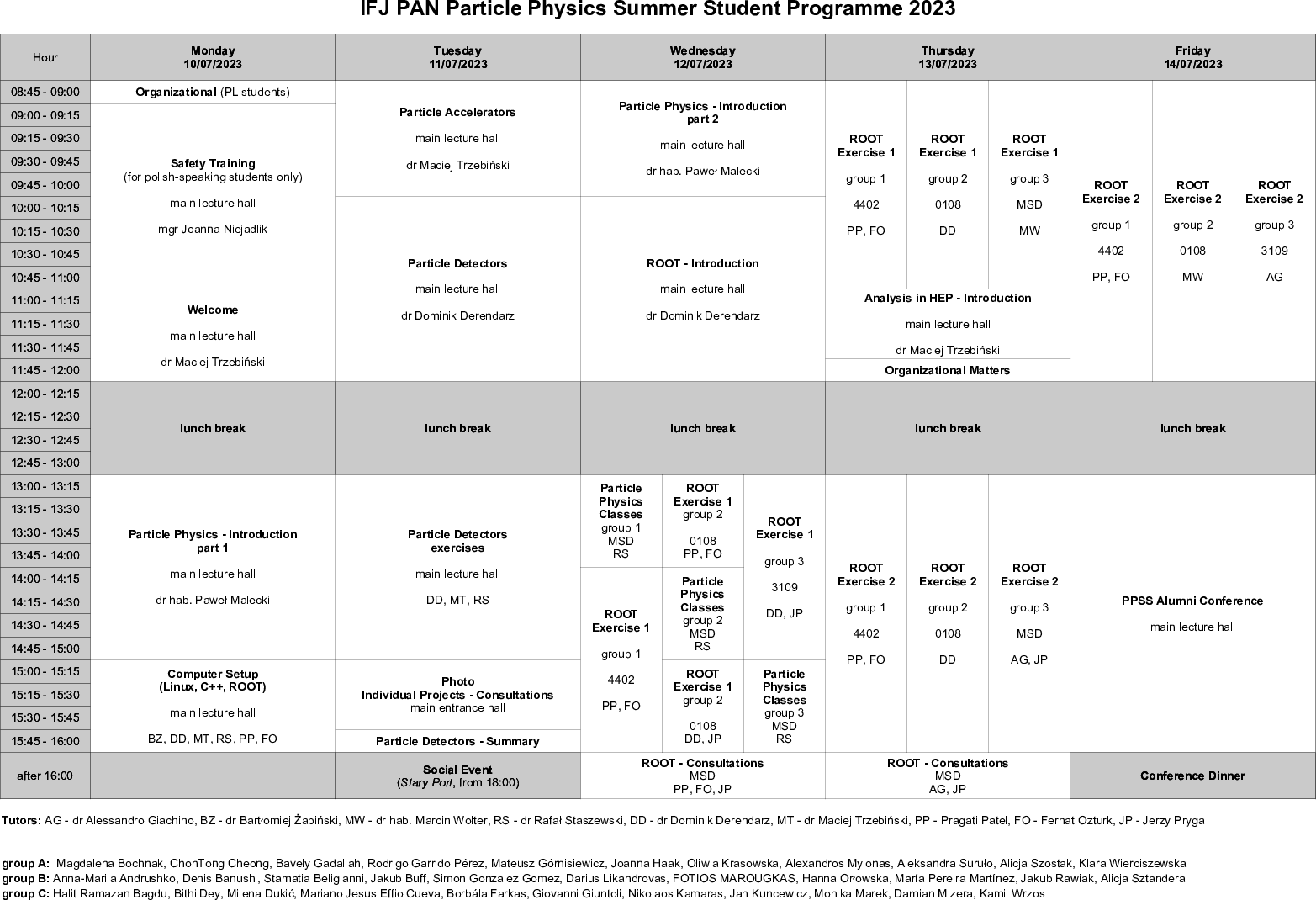}
    \caption{PPSS 2023: poster (\textbf{left}) and plan of the first week (\textbf{right}).}
    \label{PPSS2022}
\end{figure}

Traditionally, at the last day a mini-conference devoted to project reports was organized. Audience decided that the best presentation was given by Mr. Nikolaos Kamaras and Mr. Halit Ramazan Bagdu\footnote{\textit{Identification of b-jets produced in heavy-ion collisions with neural network(s)}; dr Dominik Derendarz.}. The second place was won by Ms. Aleksandra Suruło and Ms. Stamatia Beligianni\footnote{\textit{Estimation of QCD critical signatures in SHINE heavy-ion collisions through proton intermittency analysis and Monte Carlo simulations}; dr Nikolaos Davis.} and the third \textit{ex aequo} by Ms. Borbala Farkas and Mr. Alexandros Mylonas\footnote{\textit{Bayesian Analysis of Cosmic Rays data for Earthquake Predictions}; dr David Alvarez.} and Mr. Fotios Marougkas and Ms. Alicja Sztandera\footnote{\textit{Studying high-energy processes in nuclear collisions at the LHC and future colliders}; dr Richard Ruiz.}.

\newpage
\mbox{\ }
\newpage

\begin{papers}

\ResCnt

\maketitle

\begin{abstract}
Jets are copiously produced in heavy-ion collisions at the LHC energies. Their calibration is crucial for precise measurements of various processes, such as top-quark pair production. In this study, jet energy scale and resolution are evaluated in 2016 proton-lead collisions collected at centre-of-mass energy of 8.16~TeV per nucleon pair, corresponding to a total integrated luminosity of 165~nb$^{-1}$. The so-called \textit{truth method} is applied to study differences between reconstructed and generated jets in Monte Carlo simulation. The other method involving the balance between $Z$-boson and jet transverse momenta is used to estimate jet performance in both data and Monte Carlo simulation. Performance of jet reconstruction with two jet definitions, referred to as particle flow~(PFlow) and heavy ion~(HI), is evaluated and results are compared.
\end{abstract}

\section{Jet reconstruction}
\label{sec:jet}

The baseline jet reconstruction algorithm used in the ATLAS experiment~\cite{bib:atlas} at the Large Hadron Collider~(LHC) is anti-$k_t$~\cite{bib:antiKt}. In this study, jets are clustered within the radius of $R=0.4$. Two jet definitions, referred to as particle flow~(PFlow) and heavy ion~(HI), are considered. The analysis uses $Z \rightarrow \ell\ell$ events with $\ell=e^\pm,\mu^\pm$ in 2016 proton-lead (p+Pb) data and corresponding Monte Carlo~(MC) simulation.

PFlow jets~\cite{bib:PFlow} are reconstructed by clustering four vectors corresponding to a combination of measurements from the inner detector and the calorimeter. Topological clusters with low energies are replaced by track momenta matched to those clusters. Jet calibration derived for high-pileup 13~TeV proton-proton~($pp$) collisions is used.

HI jets~\cite{bib:HI} are built using massless calorimeter towers with size of $\Delta\eta \times \Delta\phi = 0.1 \times \pi/32$. The background energy originating from the underlying event is subtracted from every tower on an event-by-event basis. Jet calibration dedicated for low-pileup p+Pb collisions is applied.

\section{Truth method}
\label{sec:reco}

Jet performance can be evaluated by comparing reconstructed jets with generated ones in simulation, referred to as the \textit{truth method}~\cite{bib:PFlow}. Generated jets, available in MC simulation, are built from stable final-state particles originating from the primary vertex, excluding muons and neutrinos. Reconstructed and generated jets are geometrically matched by imposing a distance requirement, $\Delta R<0.4$. The per-event jet $p_\mathrm{T}$ response, defined as $p^\mathrm{reco}_\mathrm{T}/p^\mathrm{truth}_\mathrm{T}$, is studied in multiple jet $p^\mathrm{truth}_\mathrm{T}$ bins. $p^\mathrm{reco}_\mathrm{T}$ and $p^\mathrm{truth}_\mathrm{T}$ denote transverse momenta of the reconstructed and matched generated jet, respectively. The mean jet response $\langle p^\mathrm{reco}_\mathrm{T}/p^\mathrm{truth}_\mathrm{T} \rangle$ is obtained as the mean of a Gaussian function fitted to the jet $p_\mathrm{T}$ response distribution. The jet $p_\mathrm{T}$ resolution is estimated as the ratio of the standard deviation of the Gaussian fit over its mean.

Figure~\ref{fig:truth} shows the mean jet response and $p_\mathrm{T}$ resolution evaluated in MC simulation for the PFlow and HI jets. The mean jet response is found to be above unity for both jet definitions, which can be understood by a quark-dominated composition of $Z \rightarrow \ell\ell$ events. Significantly larger values at low $p^\mathrm{truth}_\mathrm{T}$ for the PFlow jets, which decrease with jet $p_\mathrm{T}^\mathrm{truth}$, come from the underlying-event contribution in p+Pb collisions. This effect is not observed for the HI jets, which include the underlying-event correction. The jet $p_\mathrm{T}$ resolution determines the amount of fluctuation in the jet energy reconstruction. The resolution improves with rising $p^\mathrm{truth}_\mathrm{T}$ for both PFlow and HI jets. It amounts to 20\% at low $p^\mathrm{truth}_\mathrm{T}$ and drops to 5\% for higher $p^\mathrm{truth}_\mathrm{T}$.

\begin{figure}[htb]
	\centering
	\includegraphics[width=0.48\textwidth]{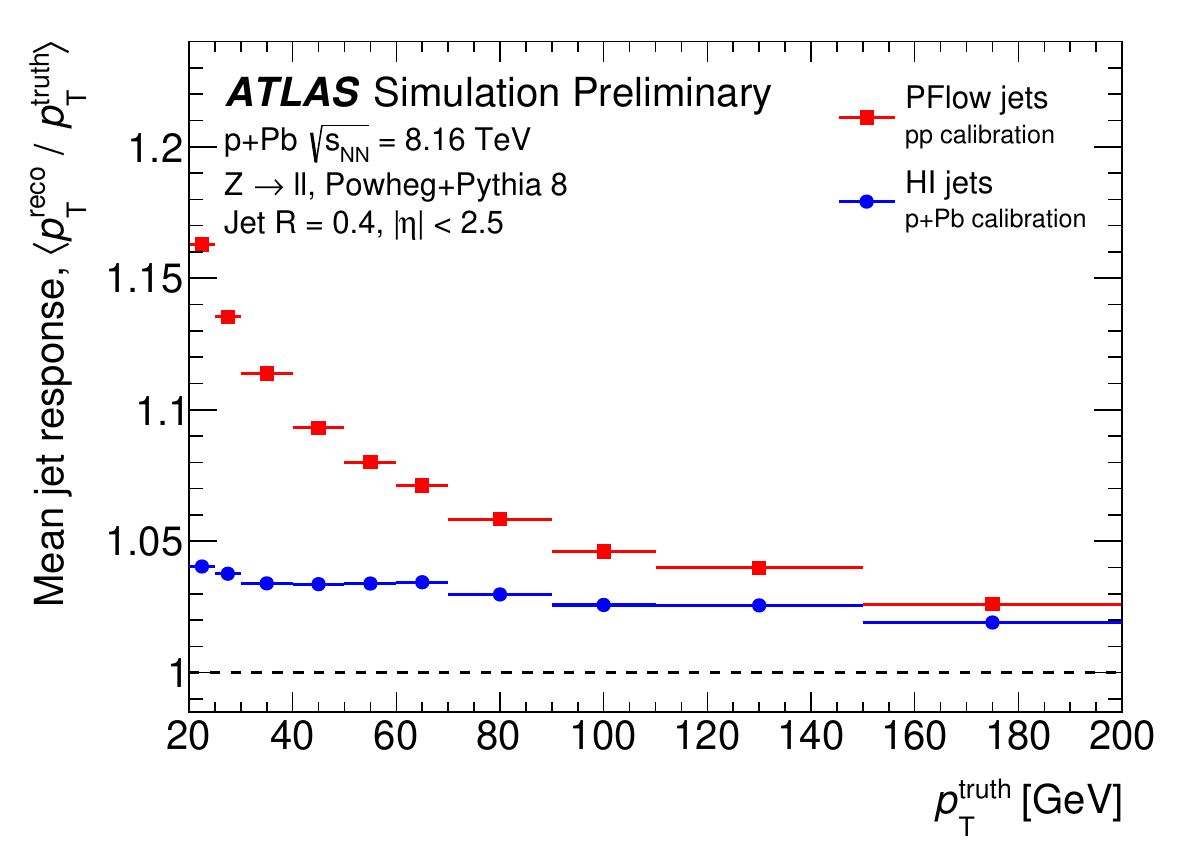}
	\includegraphics[width=0.48\textwidth]{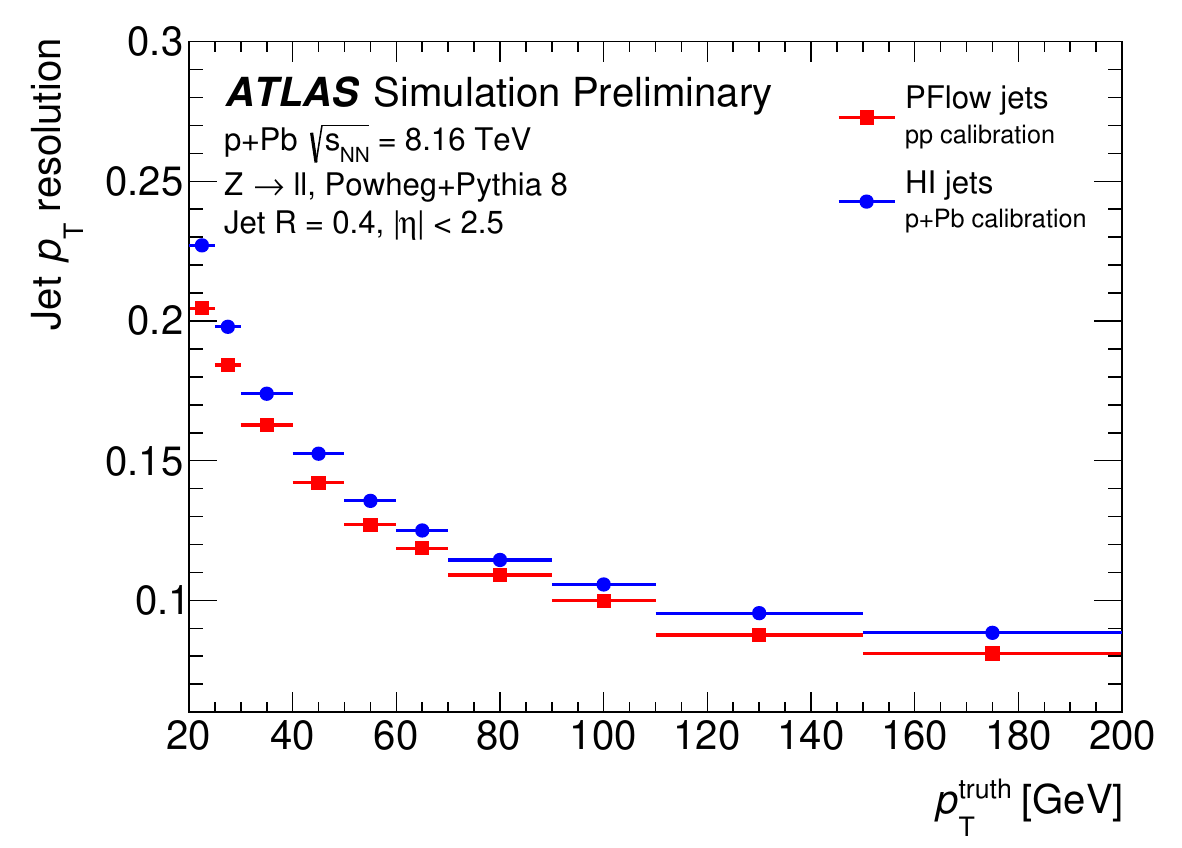}
	\caption{The mean jet response~(left) and jet $p_\mathrm{T}$ resolution~(right) in 2016 p+Pb simulation as a function of generated jet $p^\mathrm{truth}_\mathrm{T}$ for the PFlow~(squares) and HI~(circles) jets~\cite{bib:plots}.}
	\label{fig:truth}
\end{figure}

\section{Z-jet balance method}
\label{sec:ref}

Another way to estimate jet performance is based on a momentum balance between a $Z$ boson and a jet, called the \textit{Z-jet balance method}~\cite{bib:PFlow}. It utilises events with a jet recoiling against the $Z$ boson, where latter decays to either an electron or a muon pair. A pairing criterion on the azimuthal angle between the $Z$ boson and the jet, $|\Delta\phi(\mathrm{Z,jet})|>2.8$, is imposed to ensure the back-to-back emission. In this method, the jet $p_\mathrm{T}$ response is determined as $p^\mathrm{reco}_\mathrm{T}/p^\mathrm{ref}_\mathrm{T}$, where the reference transverse momentum $p^\mathrm{ref}_\mathrm{T} = p^\mathrm{Z}_\mathrm{T} |\cos \Delta\phi(\mathrm{Z},\mathrm{jet})|$ is the projection of the $Z$ boson transverse momentum $p^\mathrm{Z}_\mathrm{T}$ along the jet axis. The mean jet response $\langle p^\mathrm{reco}_\mathrm{T}/p^\mathrm{ref}_\mathrm{T} \rangle$ is defined as the mean of a Gaussian function fitted to the jet $p_\mathrm{T}$ response distribution, while the jet $p_\mathrm{T}$ resolution is obtained as the ratio of the standard deviation of the same fit over its mean.

Figure~\ref{fig:ref_pFlow} presents the mean jet response and $p_\mathrm{T}$ resolution in data and MC simulation for PFlow jets. The mean jet response is below unity over entire $p^\mathrm{ref}_\mathrm{T}$ range. The response at the lowest $p^\mathrm{ref}_\mathrm{T}$ is increased due to the underlying-event contribution. The resolution improves with rising $p^\mathrm{ref}_\mathrm{T}$. The overall jet $p_\mathrm{T}$ resolution is higher compared to the truth method due to intrinsic broadening coming from physics of $Z \rightarrow \ell\ell$ decays. A good agreement is found between data and MC simulation within systematic uncertainties  in the mean jet response, while a small MC non-closure is observed in the jet $p_\mathrm{T}$ resolution.
\vspace*{-0.2cm}
\begin{figure}[htb]
	\centering
	\includegraphics[width=0.35\textwidth]{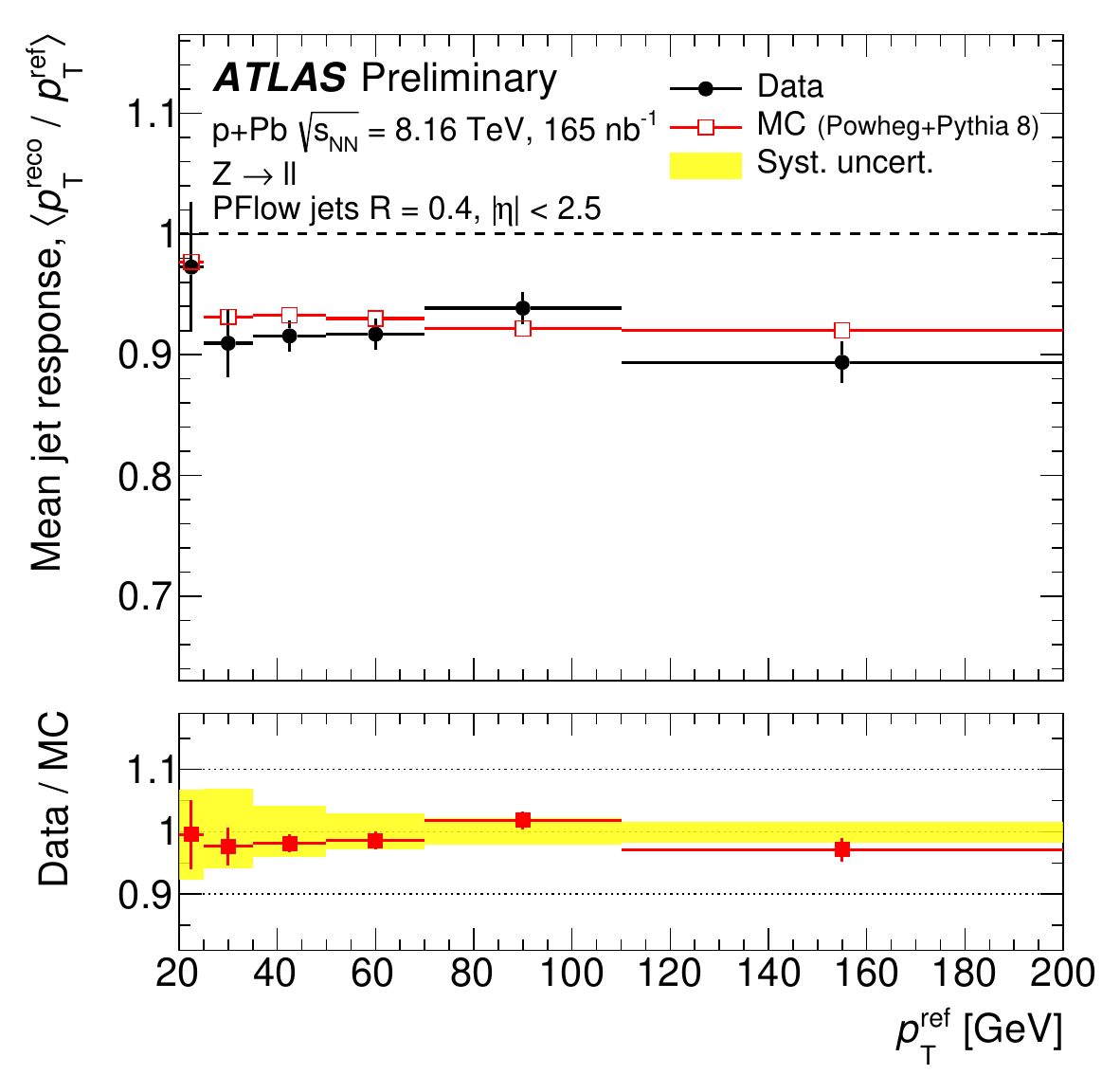} \hspace{0.08\textwidth}
	\includegraphics[width=0.35\textwidth]{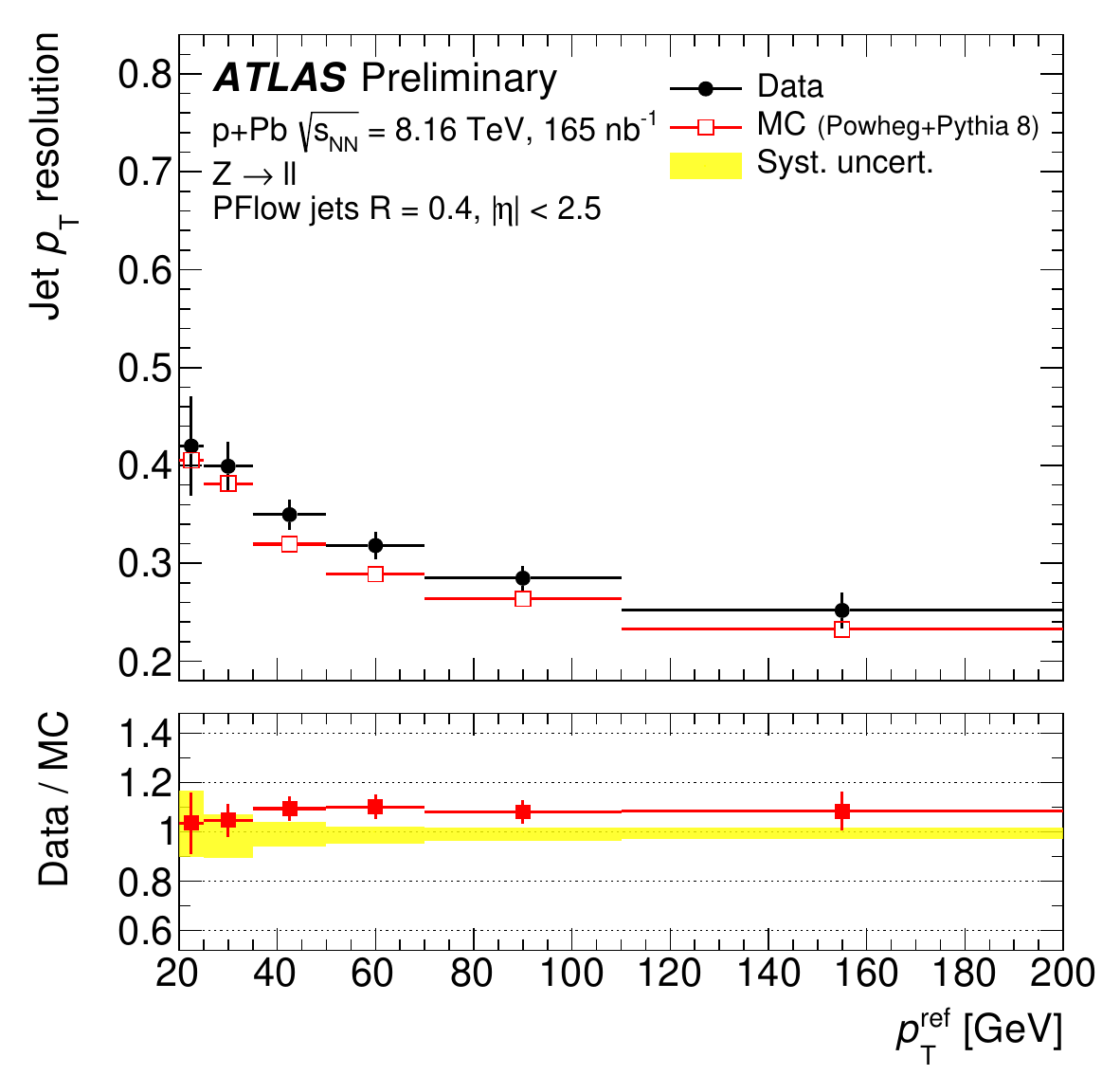}
	\caption{The mean jet response~(left) and $p_\mathrm{T}$ resolution~(right) in 2016 p+Pb data and simulation as a function of reference jet $p^\mathrm{ref}_\mathrm{T}$ for PFlow jets~\cite{bib:plots}.}
	\label{fig:ref_pFlow}
\end{figure}
\vspace*{-0.2cm}
The mean jet response and $p_\mathrm{T}$ resolution in 2016 p+Pb data and MC simulation for HI jets are shown in Figure~\ref{fig:ref_HI}. The mean jet response is below unity and rises with $p^\mathrm{ref}_\mathrm{T}$ as expected. The resolution at low $p^\mathrm{ref}_\mathrm{T}$ is worse compared to PFlow jets. The mean jet response in data and MC simulation is consistent within uncertainties, while a small MC non-closure is found in the jet $p_\mathrm{T}$ resolution at jet $p_\mathrm{T}>50$~GeV.

\begin{figure}[htb]
	\centering
	\includegraphics[width=0.35\textwidth]{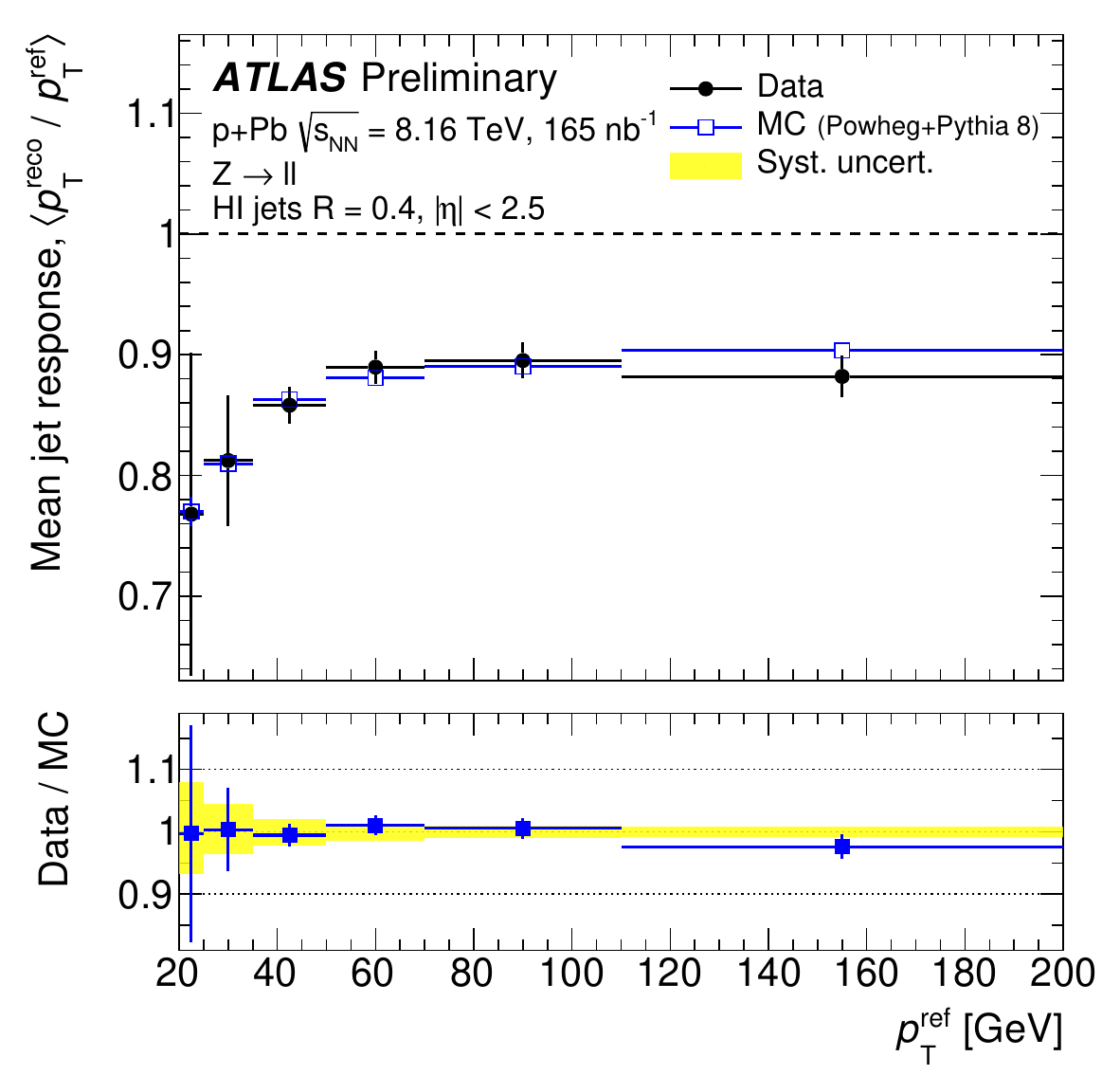} \hspace{0.08\textwidth}
	\includegraphics[width=0.35\textwidth]{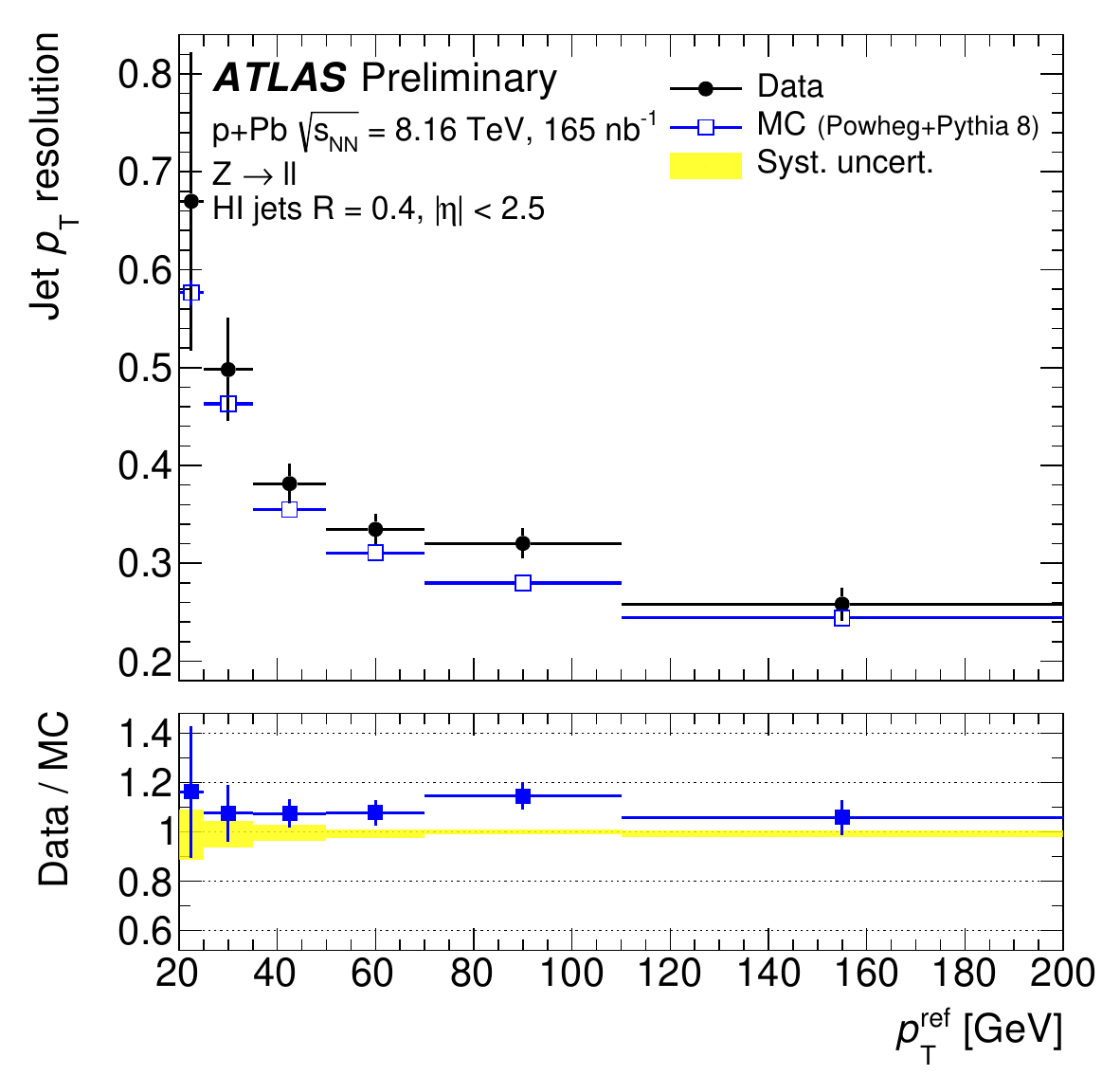}
	\caption{The mean jet response~(left) and $p_\mathrm{T}$ resolution~(right) in 2016 p+Pb data and simulation as a function of reference jet $p^\mathrm{ref}_\mathrm{T}$ for HI jets~\cite{bib:plots}.}
	\label{fig:ref_HI}
\end{figure}

\FloatBarrier
\section{Conclusion}
\label{sec:conclusion}

The jet performance has been evaluated in 2016 p+Pb collisions at $\sqrt{s_{\mathrm{NN}}}=8.16$~TeV collected by ATLAS. Higher mean jet response at low jet $p_\mathrm{T}$ for PFlow jets originates from the underlying-event contribution in p+Pb collisions. The resolution improves with increasing jet $p_\mathrm{T}$ for both jet definitions. The results have been obtained in data and MC simulation with systematic uncertainties. A reasonable agreement in the mean jet response and $p_\mathrm{T}$ resolution is observed for both jet definitions. The studies are a key input to the analysis of top-quark pair production in p+Pb collisions~\cite{bib:ttbar}.
\vspace*{-0.2cm}
\section*{Acknowledgements}
\label{sec:acknowledgements}

This work was partly supported by the National Science Centre of Poland under the grant number 2020/37/B/ST2/01043 and PL-Grid Infrastructure.


\ResCnt

\maketitle

\vspace*{-1.2cm}
\begin{center}
University of Warsaw
\end{center}

\begin{abstract}
Giant Radio Array for Neutrino Detection (GRAND) is a planned large scale detector aimed at detection of Ultra High Energy cosmic rays, neutrinos and photons using radio emission of extensive air showers. It is currently in prototype stage. In its final form it will consist of 200 000 radio antennas covering 200 000 km$^2$ of land and will have the potential to discover the first ever UHE neutrinos. I will discuss detector design, goals, challenges and timeline of the experiment.
\end{abstract}

\section{Introduction}
Ultra High Energy Cosmic Rays (UHECR) are charged particles of cosmic origin, carrying energy greater that 1 EeV(=$10^{18}$ eV). The existence of UHECR is experimentally confirmed, but UHE photons and neutrinos remain to be discovered. Since photons and neutrinos are produced in the interaction of UHECR with the Cosmic Microwave Background (CMB), their existence is expected. The Giant Radio Array for Neutrino Detection (GRAND) aims at discovering those particles for the first time in history.

Figure \ref{fig:merged} (left) shows the measured energy spectrum of UHECR and the predicted energy spectrum of cosmogenic photons and neutrinos. The red circle represents the GRAND target spectrum -- the experiment is predicted to measure cosmogenic photons and neutrinos even for pessimistic (low flux) scenarios. Moreover, GRAND will be able to detect other astrophysical phenomena, such as Giant Radio Pulses and Fast Radio Bursts.

\begin{figure}[!htbp]
    \centering
    \includegraphics[width=\textwidth]{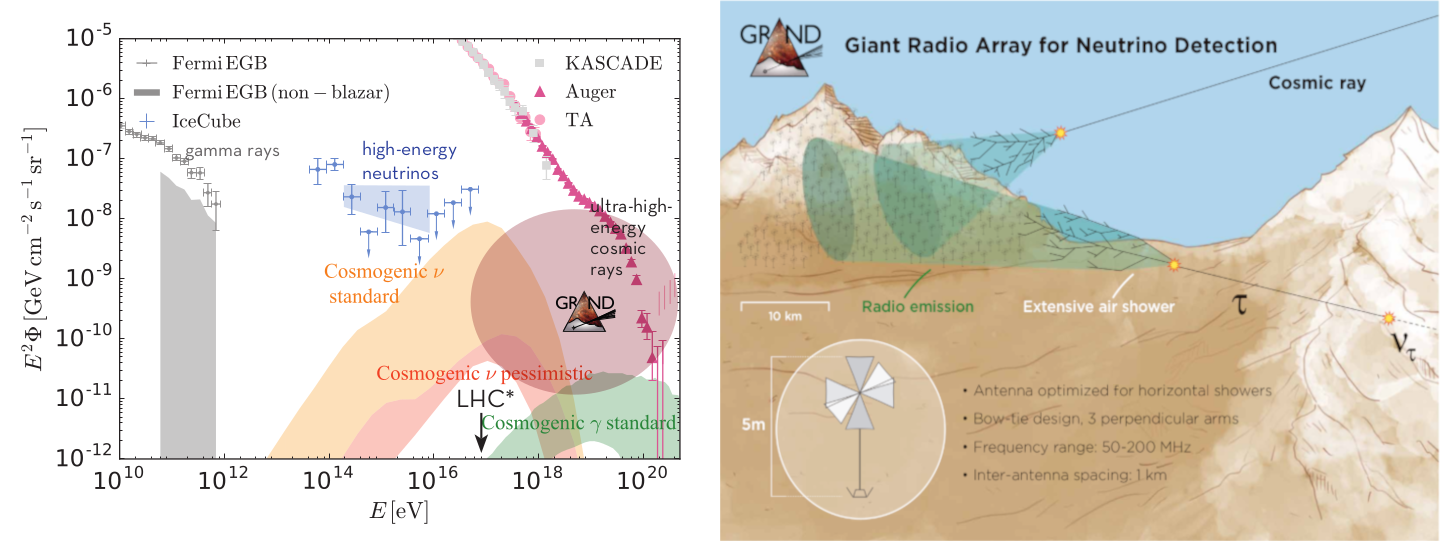}
    \caption{Energy spectrum range of GRAND (red circle), measured flux of UHECR and predicted flux of cosmogenic photons and neutrinos (left); A schematic visualization of the working principle of GRAND (right) \cite{grand}}
    \label{fig:merged}
\end{figure}

\section{The working principle}
    \subsection{Extensive Air Showers}
Extensive Air Showers (EAS) are cascades of particles caused by an interaction of a particle incoming from space with an air molecule. The cascade development produces an electromagnetic wave in radio frequency. The following are two mechanisms contributing to the emission of this signal \cite{air-showers}:
\begin{description}
    \item[Charge separation] The EAS contains particles of opposite charges moving in the same direction. In the presence of Earth's magnetic field, oppositely charged particles will bend in opposite directions, creating a current that induces the radio pulse.
    
    \item[Negative charge excess (Askaryan effect)]  Due to Compton scattering, the EAS carries an excess of electrons. This implies an electric current along the shower propagation direction and induction of an electromagnetic radio signal.
\end{description}

EAS emit signal in the direction of their propagation, in a narrow cone (1-2$^\circ$). The signal has the highest amplitude around the edges of the cone, creating a pattern similar to Cherenkov rings.

    \subsection{Neutrino detection}
An interaction of neutrino with air, which is a very thin medium, has a very small cross section. In rock, however, the interaction length is of the order of hundreds of km for neutrino energy $\sim1$ EeV \cite{grand} \cite{neutrinos}. A neutrino can approach Earth from a nearly horizontal direction, skim the Earth surface and produce (in a charged current interaction) a lepton of a corresponding flavor. Let's consider all 3 flavors of incoming neutrinos:
\begin{description}
    \item[Electron neutrino] An electron will be produced and absorbed inside the rock. The EAS will not occur.
    
    \item[Muon neutrino] A muon will be produced inside the rock, travel through it and escape into the atmosphere. Due to its large lifetime it will decay and start an EAS too far from the detector
    
    \item[Tau neutrino] A tau will be produced inside the rock, travel through it and escape into the atmosphere. Due to its short enough lifetime it will decay and start an EAS in the range of the detector.
\end{description}
In principle, GRAND is sensitive only to tau neutrinos, which are not produced in nucleus-CMB interaction, but neutrino oscillations guarantee a significant flux of cosmogenic tau neutrinos. Figure \ref{fig:merged} (right) shows a schema of the detector and its detection principle.

\section{Detector}
    \subsection{Design}
GRAND, in its final form will consist of 20 sub-arrays of 10 000 radio antennas each. The antennas will be spaced 1 km apart, creating a sparse array and covering of 200 000 km$^2$ area. The detector will work in 50-200 MHz frequency band and will be sensitive to air showers coming from 85-95$^\circ$ zenith angle. This refers to EAS started by nuclei and photons too, because the radio footprint of vertically incoming EAS is too small to be registered by such a sparse array. The preferred locations for the antenna arrays are mountain slopes, because EAS started by neutrinos may propagate upwards (Fig. \ref{fig:merged}).
    \subsection{Efficiency}
One of the qualities of GRAND will be an excellent angular resolution (better than 0.2$^\circ$). That could allow to locate sources of UHE neutrinos and photons (charged particles are not fit for this purpose, as they are deflected by magnetic fields). The energy resolution will reach about 15\%. Due to Earth's rotation, the detector will cover around 80\% of sky each day. Identification of incoming particles will be possible, using atmospheric depth of the shower maximum.
    \subsection{Background}
Some of the background sources are man-made, such as FM radio stations, noise from the electronic components, power lanes, planes etc. It is possible to avoid some of them by placing the detector arrays far from populated areas. There's also natural background e.g. radiation from the galactic plane, storms. In principle, EAS radio emission has a characteristic signature, so high efficiency of background rejection will be achievable.

\section{Outlook}
 GRAND is a promising new experiment, currently in prototype stage. At the next stage of development -- GRANDProto300 -- it will consist of 300 antennas (possibly paired with scintillator detectors) and will be able to detect very inclined cosmic rays. The first sub-array, GRAND10k (2025) could possibly detect first ever cosmogenic neutrinos, if their flux is high enough. GRAND200k (2030s) will achieve unprecedented sensitivity and will have potential to discover cosmogenic photons and neutrinos, even for a pessimistic flux predictions

\ResCnt

\maketitle
\vspace*{-1.2cm}
\begin{center}
Bachelor's thesis research at University of Cyprus
\end{center}

\begin{abstract}

Silicon particles in an evanescent wave trap create one-dimensional structures through optical binding. As particle size increases, so does the separation between particles. The spring constant increases with laser power and particle size due to stronger binding forces and radiation pressure.
\end{abstract}

\section{Introduction}

The concept behind one dimensional optical trapping using evanescent field  is that a light beam can manipulate and organize dielectric materials into a chain (Fig. \ref{fig1}), such as silicon particles in ionized water.

\begin{wrapfigure}{l}{0.40\textwidth}
    \centering
    \includegraphics[width=0.40\textwidth]{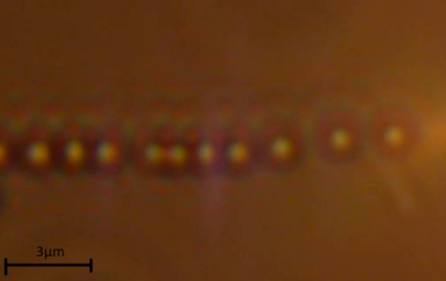}
    \caption{Self-organised 1 $\mu$m silicon particles in an evanescent field trap with 375mW laser power forces.}
    \label{fig1}
\end{wrapfigure}

The key forces at play in this experiment are the gradient force, scattering force, and optical binding force, determining the interaction strength within the bound chains. In this thesis experiment, radiation pressure compels particles to move in the response to scattered light, achieved through the use of two counter-strike beams for balance. The optical binding force is what organizes particles into chains, with each particle scattering laser light and influencing nearby particles. These particles act as lenses, concentrating scattered light into the ''hot spots'', the location of which depends on particle properties and the electromagnetic field's morphology \cite{2}.

\section{Experimental Setup}

As shown in Fig. \ref{fig2}, a  high-power white LED array (MWWHL3 by Thorlabs Inc.) serves as the illumination source, directed by a 45-degree silver mirror mounted on a cube. A CMOS camera (CS165CU by Thorlabs) captures scattered light from the particles, while an infrared cut-off filter eliminates scattered laser light for clear imaging.

The glass prism's upper surface is covered with a silica microparticles placed in ionized water and sealed with a microscope cover slip. Two silicon particle sizes, 1 $\mu$m and 2 $\mu$m in diameter, were used.
\begin{wrapfigure}{l}{0.65\textwidth}
    \centering
    \includegraphics[width=0.65\textwidth]{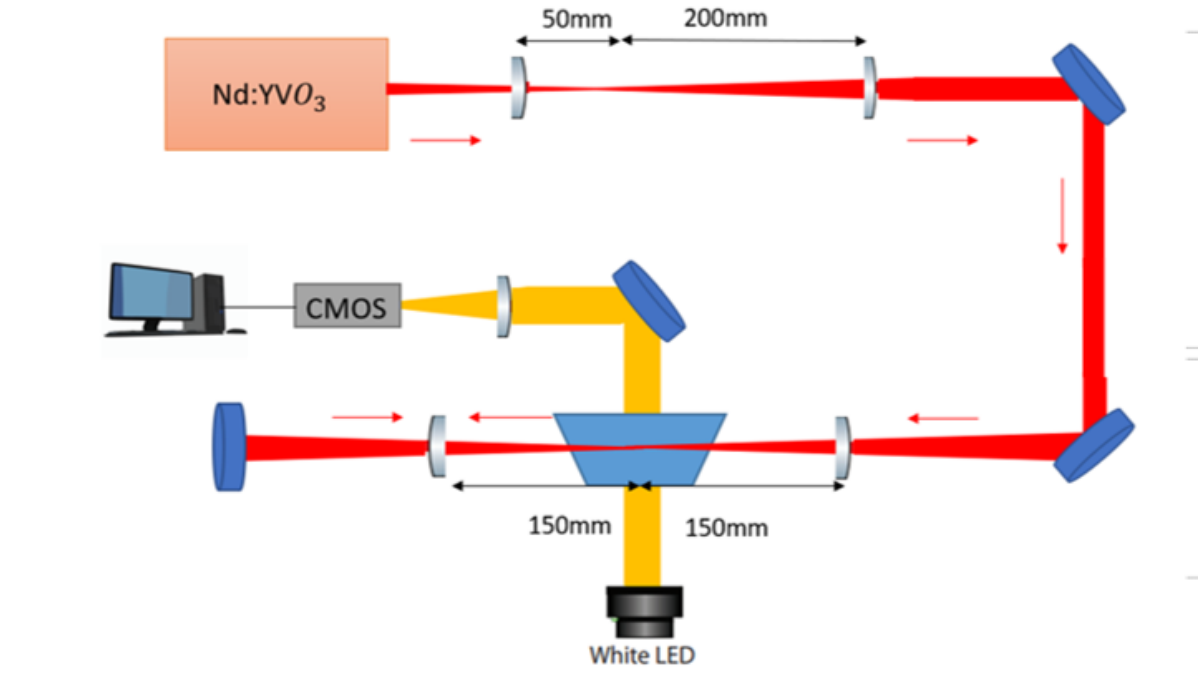}
    \caption{Schematic of the experimental set-up used for optical
binding experiment.}
    \label{fig2}
\end{wrapfigure}

An Nd:YVO$_3$ laser system at 1064 nm wavelength and a max output power of 1 W (MIL III 1064 1W by CNI LASER) is the laser source. The laser beam passes through a telescope arrangement with 50 mm and 200 mm lenses, expanding into an 8-mm collimated beam. Afterward, the beam encounters an N-BK7 Dove prism (PS991 by Thorlabs), creating an evanescent field through the total internal reflection. To balance the radiation pressure, the reflected beam strikes a retro-reflecting mirror and symmetrically re-enters the prism at a critical angle from the opposite side.

A free video modeling and analysis software called \textsc{Tracker} \cite{3} used to analyse the dynamics of the optically bound particles. \textsc{Tracker} returns the $x$ and $y$ position of a center of mass vs time step of a particle.

\section{Harmonic Oscillation of Particles}
 It can be considered that a particle in a chain is moving inside a harmonic potential. Having this in mind and combining some basic thermodynamics equations the following equation can be obtained:
\begin{equation}
    \frac{1}{2}K_{B}T=\frac{1}{2}\kappa\langle x \rangle ^2.
    \label{eq1}
\end{equation}

In this equation, \emph{$K_{B}$} is the Boltzmann constant, $T$ is temperature, \emph{$\kappa$} is the spring constant, and $\langle x \rangle$ is the average displacement. The equation connects mean energy in thermodynamics (left) to the potential energy in a harmonic oscillator (right). The experiment aims to evaluate particle chain stability by calculating the spring constant using a Gaussian equation for one-dimensional ($x$) particle motion:
\begin{equation}
    f=Ae^{\frac{-\langle x \rangle ^2}{2w^2}},
    \label{eq2}
\end{equation}
where $w$ is the width of Gaussian fit. By combing Eq. \ref{eq1} and \ref{eq2}, the spring constant can be calculated :
\begin{equation}
    \kappa=\frac{K_{B}T}{2w^2}.
    \label{eq3}
\end{equation}

\section{Results}

With the data extracted from the \textsc{Tracker}, we plot the Gaussian histogram of each particle's $x$ position of a specific size at a specific laser power and find the average value. The following graphs Fig.\ref{fig3} Fig.\ref{fig4} are the result of averages values.

\begin{figure}[!htbp]
  \centering
    \includegraphics[width=0.49\textwidth]{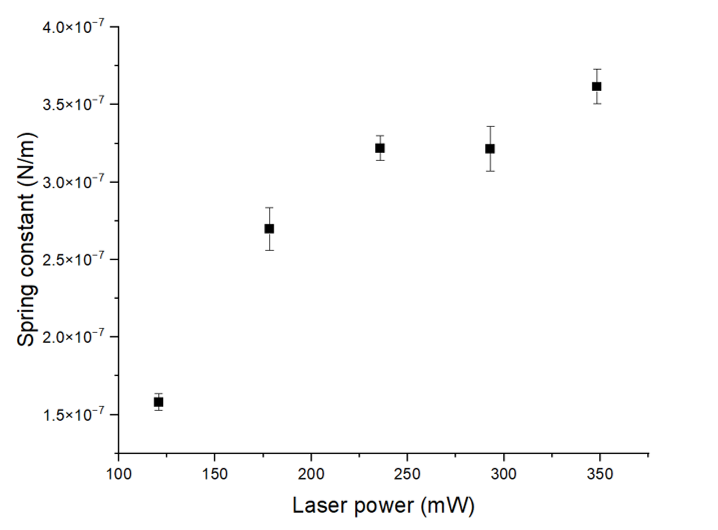}
    \includegraphics[width=0.49\textwidth]{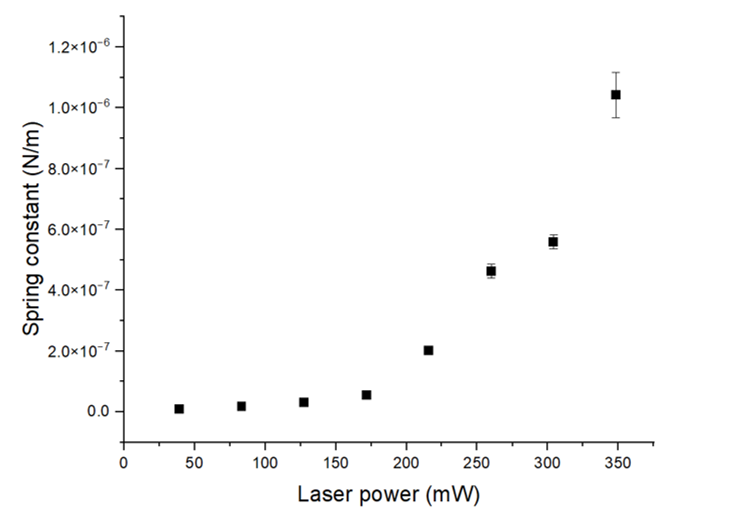}
    \caption{Spring constant of 1 $\mu$m (left) and 2 $\mu$m (right) silica spheres as a function of the laser power.}
    \label{fig3}
\end{figure}

In Fig. \ref{fig3} the spring constant is plotted against laser power for 1 $\mu$m and 2 $\mu$m particles respectively. As expected, the spring constant increases with power for all sizes due to intensified optical forces maintaining particle formations. Notably, 2 $\mu$m particles require less power for trapping than 1 $\mu$m particles, attributed to the reduced Brownian motion. Additionally, regardless of the particle size, increasing laser power enhances the spring constant by strengthening the optical binding force, resulting in greater particle stability due to increased total confining force intensity.

\begin{figure}[!htbp]
  \centering
    \includegraphics[width=0.49\textwidth]{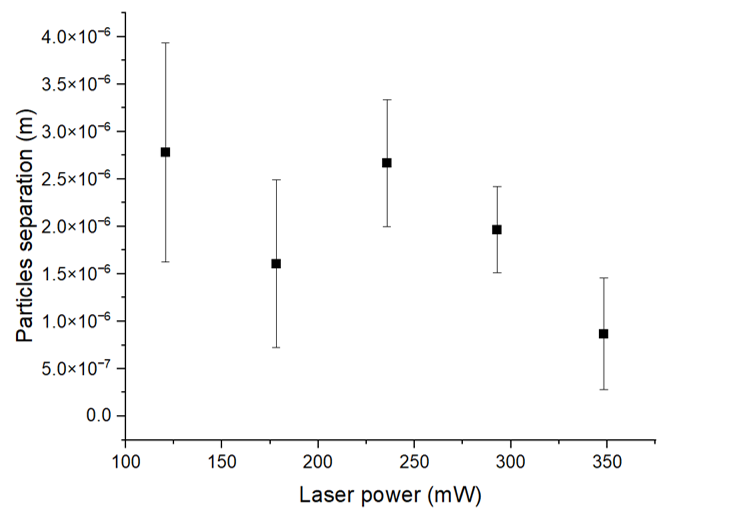}
    \includegraphics[width=0.49\textwidth]{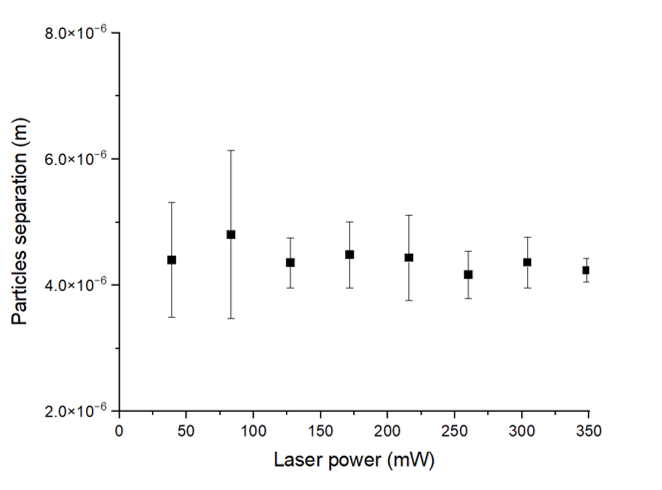}
    \caption{Inter-particle separation of 1 $\mu$m (left) and 2 $\mu$m (right) silica spheres as a function of laser power.}
    \label{fig4}
\end{figure}

As illustrated in Fig. \ref{fig4}, particle separation remains constant regardless of changes in the laser power. Instead, particle separation is dependent on the particle size. For 2 $\mu$m particles, the separation was measured at d=4.29$\pm$0.13 $\mu$m, while for 1 $\mu$m silica spheres, the averaged separation was d=2.43$\pm$0.33 $\mu$m. It is evident that as the particle diameter increases, so does the particle separation. These experimental results suggest that the optical binding force is the most significant factor in our experiment. Larger particles create a ``hot'' spot at a different distance from their center, resulting in trapping sites at different positions compared to smaller particles. This explains why larger particles exhibit greater separation between them.

\ResCnt

\maketitle

\begin{abstract}

CPT symmetry is assumed to be strictly conserved in the Standard Model (SM). Consequently, detection of even a minor deviation from CPT invariance would be hinting at a more fundamental theory, possibly at the Planck energy scale. Current technology enables us to explore energies nearing the Planck scale by probing space-time symmetry violations. The framework to study these deviations is called the Standard Model Extension (SME). This paper includes the formalism used for testing CPT with neutral $D$ mesons and a discussion on how to extract the most stringent limits on SME CPTV parameters in the charm sector using LHCb data. 

\end{abstract}

\section{Introduction}

Flavoured neutral mesons are neutrally charged mesons that have non-zero strangeness, charm or beauty content. For such mesons, weak interactions lead to spontaneous transitions between their matter and antimatter states, which are called oscillations. Consequently, meson and antimeson states (flavour states) \footnote{$\ket{D^{0}}$,$\ket{\bar{D}^{0}}$ for neutral D meson} are not the eigenstates of the Hamiltonian. The eigenstates of the Hamiltonian, called mass states, can be linearly expressed in terms of flavour states. The frequency of neutral meson oscillations is effectively controlled by the difference of energies between mass states. Meson and antimeson will interact differently with the hypothetical CPTV/LV field which in turn will affect the aforementioned energy difference. Despite the small scale of these hypothetical effects, they should be observable through oscillation methods \cite{kostelecky}. More specifically, we can test CPTV and LV by analysing the time-dependent asymmetry constructed from decay rates of $D^{0}$ and $\bar{D}^{0}$ into final states that preserve the flavour of the initial state: $f=K^{-}\pi^{+}$, $\bar{f}=K^{+}\pi^{-}$. The asymmetry constructed from such decay rates is called right-sign. To measure the couplings of meson valence quarks to CPTV/LV fields\footnote{quark couplings to CPTV/LV fields are SME CPTV target observable}, we can analyse this asymmetry in various ways. This includes examining the asymmetry in both decay time and sidereal time, as well as, considering different meson momentum bins.

\section{Neutral D meson system}

A neutral D meson system is expressed as a linear combination of the Schrodinger wave functions for $D^{0}$ and $\bar{D}^{0}$. The time evolution of the meson system is governed by the Schrodinger-type equation with an effective Hamiltonian $H_{\text{eff}}$. CPT symmetry is conserved in the neutral meson system when the difference of the diagonal elements of $H_{\text{eff}}$ is zero. For the purposes of CPT studies with neutral mesons it is convenient to define a complex CPTV parameter: $z=\frac{H^{11}_{\text{eff}}-H^{22}_{\text{eff}}}{\Gamma(x-iy)}$\footnote{$x,y$ are parameters controlling the frequency and shape of oscillations, $\Gamma$ $D^{0}$ decay width. \cite{focus, hflav}}.
If we follow the procedure described in \cite{kostelecky} we will be able calculate the time evolution of $D^{0}$ and $\bar{D}^{0}$ that can subsequently be used to obtain probability densities: $P_{f}(t)=|\bra{f}T\ket{D^{0}(t)}|^{2}$, $\bar{P}_{\bar{f}}=|\bra{\bar{f}}T\ket{\bar{D}^{0}(t)}|^{2}$, where $f=K^{-}\pi^{+}$,~$t$~denotes the decay time of meson. Using the probability densities introduced above the right-sign asymmetry can be formed:

\begin{equation}
A_{CPT}(t)=\frac{\bar{P}_{\bar{f}}(t)-P_{f}(t)}{\bar{P}_{\bar{f}}(t)+P_{f}(t)}.
\end{equation}

According to \cite{focus} the above expression can be approximated via Taylor expansion to the 3rd degree in $x,y$ under the assumption that CPTV is small $|z|^{2}<<1$ \footnote{$\delta$, $\phi$ represent strong and weak phase. eigenvalues~\cite{focus, hflav}.}:

\begin{equation}
\begin{aligned}
A_{CPT}(t) = & \left( y \text{Re}(z) - x \text{Im}(z)\right)t - \sqrt{R} \sin{\phi} \left(x \cos{\delta} - y \sin{\delta}\right)t \\
& - \text{Re}(z) \cos{\phi} \left( \frac{\sqrt{R}(x^2 + y^2)   (x \cos{\delta} - y \sin{\delta})}{2x} \right)t^2 \\
& + \frac{ \text{Re}(z) }{6} t^3 x^2 y + \frac{ \text{Re}(z) }{6} t^3 y^3.
\label{simplifed}
\end{aligned}
\end{equation}

In experimental tests of CPT symmetry with neutral mesons the observable is formed from the right-sign asymmetry of measured $D^{0}$ decay rates. CPTV parameters can then be extracted from the fit of (\ref{diff}) to the observable. 

\section{Standard Model Extension}

In SME CPTV/LV effects are treated as small deviations to the SM. 
To find an expression for parameter $z$ in SME we need to derive the difference
of the diagonal terms of $H_{\text{eff}}$ \cite{kostelecky}. This leads to the following expression for $z$:

\begin{equation}
z \approx \frac{\beta^{\mu}\Delta a_{\mu}}{\Gamma \left(x-i y\right)},
\end{equation}

\noindent where $\beta^{\mu}=\gamma(1,\vec{\beta})$ represents four-velocity in the observer frame. $\Delta a_{\mu}$ are CPTV parameters connected with coefficients ($a^{q_{1}}_{\mu}$,$a^{q_{2}}_{\mu}$) related to the coupling of the two valence quarks with the hypothetical CPTV/LV field through relation: $\Delta a_{\mu} = r_{q_{1}} a^{q_{1}}_{\mu} - r_{q_{2}} a^{q_{2}}_{\mu}$, where $r_{q_{1}},r_{q_{2}}$ are scaling factors. The most convenient reference frame for CPTV studies is the fixed stars reference frame where $z$ is subject to sidereal modulations:

\begin{equation}
\beta^{\mu}\Delta a_{\mu} = \gamma \left[\Delta a_{0}+\beta \Delta a_{Z} \cos{\chi} + \beta \sin{\chi} \left(\Delta a_{Y}\sin{\Omega T}+\Delta a_{X}\cos{\Omega T}\right)\right],
\end{equation}

\noindent where $T$ denotes sidereal time\footnote{$\chi$ is connected with the geographical location of the lab on Earth, $\Omega$ sidereal frequency}. Consequently in SME $A_{CPT}$ will be a two-dimensional function of both decay and sidereal time ($t$,$T$).

\section{CPT parameter extraction in LHCb}

LHCb has the highest statistics in $D^{0}\rightarrow K^{-}\pi^{+}$ channel. For LHCb Run 2 (data-taking period 2015-2018) alone it is $10^{4}$ greater than the one that was used by FOCUS collaboration to extract the current best bounds on CPTV parameters in the charm sector \cite{focus}. After completion of the present data-taking period (LHCb Run 3) the available statistics should increase tenfold. To enhance the sensitivity of our CPTV measurement when dealing with a substantial CPV effect and its uncertainty, we need to modify our observable. This can be done by computing the differences in $A_{CPT}$ within sidereal time bins. Since CPV is not influenced by sidereal time, this modification effectively cancels out the CPV contribution from our dataset. If we then express parameter z in terms of SME $\Delta a_{\mu}$ coefficients, the difference of simplified expressions for right-sign asymmetry from (\ref{simplifed}) will take the form:

\begin{align}
\begin{split}
\text{Diff}(t,T_{i},T_{j}) &= A_{CPT}(t,T_{i}) - A_{CPT}(t,T_{j}) \\
&= \frac{1}{3 \Gamma} t^2 \beta \gamma (t x y + 3 \sqrt{R} (x \cos(\delta) - y \sin(\delta)) \\
&\quad \times \sin(\chi) \sin\left(\frac{1}{2}(T_{i} - T_{j}) \Omega\right) \\
&\quad \times \left(\Delta a_{Y} \cos\left(\frac{1}{2}(T_{i} + T_{j}) \Omega\right) - \Delta a_{X} \sin\left(\frac{1}{2}(T_{1} + T_{2}) \Omega\right)\right).
\end{split}
\label{diff}
\end{align}

This method offers access to two out of four SME CPTV coefficients related to the coupling of the quarks with the LV field. In the table below the result of a fit of (\ref{diff}) to toy MC right-sign asymmetry in $D^{0}\rightarrow K^{-}\pi^{+}$ mode were included. 
In the toy MC simulation, CPTV $\Delta a_{\mu}$ coefficient values were set to match the upper limit on $\Delta a_{\mu}$ achieved by FOCUS \cite{focus}. 

\begin{table}[!h]
\centering
\begin{tabular}{c|c|c}
  & $\Delta a_{X}$ [GeV] & $\Delta a_{Y}$ [GeV] \\
  \hline
  \textcolor{red}{fit} & $(4.7 \pm 0.3) \times 10^{-13}$ & $(3.5 \pm 2.9) \times 10^{-14}$ \\
  \hline
  \textcolor{blue}{MC} & $5 \times 10^{-13}$ & $0$ \\
  \hline
\end{tabular}
\end{table}

\section{Summary}
We can study CPTV and LV with flavoured neutral mesons. LHCb Run2 and Run3 have the highest statistics in $D^{0}\rightarrow K^{-}\pi^{+}$ mode. Sidereal analysis of asymmetry differences allows us to extract bounds on $\Delta a_{\mu}$. If we approximate $A_{CPT}$ by (\ref{simplifed}) we can demonstrate that the errors on $\Delta a_{\mu}$ in LHCb Run2 should be an order of magnitude smaller than those in the current most precise measurement~\cite{focus}.

\ResCnt

\maketitle

\begin{abstract}
This paper shows the evidence of Exclusive Jet production (EXC~JJ) in data collected with the ATLAS and ATLAS Forward Proton (AFP) detectors in 2017. After short introduction to experimental setup the trigger and event selection is discussed. Applied cuts show presence of EXC JJ production in one sample from all considered one.
\end{abstract}
\section{Introduction}

Located at the Large Hadron Collider (LHC) \cite{LHC}, the ATLAS experiment \cite{ATLAS} has been designed with the goal of measuring the products of proton--proton collisions. Although it has a full azimuthal angle coverage and a large acceptance in pseudorapidity, it is not fully sufficient for a certain group of physics processes, namely the diffractive physics. These processes can be characterised by the presence of the following observables: rapidity gap (a space in rapidity where no particles are produced) and protons scattered at at very small angles. 



In order to measure forward protons additional forward detectors far away from the interaction point. Such detectors are ATLAS Forward Proton (AFP) \cite{AFP} detectors with two stations~\footnote{Named Near and Far stations, where Near stations are closer to the ATLAS detectors.} located on each side of ATLAS~\footnote{Those sides are commonly named side A and C.} hundreds of meters away from interaction point.

In this paper the interest is in exclusive jet production with signatures are two jets and two forward protons. No other particles should be present in the event and two rapidity gaps are expected between protons and jets.

\section{Runs and Triggers of Interests}

In order to secure the survival of mentioned previously rapidity gap only runs (the periods of data-taking) with very small pile-up were considered in the analysis. Those were the runs with either $\mu\sim$ 1.0 (run 331020 and part of run 341649), $\mu\sim$ 0.04 (run 336505) or $\mu\sim$ 2.0 (runs 341294, 341312, 341419, 341534, 341615 and part of run 341649).

The first step of event selection starts already during data taking when very complex trigger algorithms select events, based on chosen characteristic, to be saved for various analysis. The characteristics will differ depending on studies and some of them can be determined right after the collision based on partial detector data. Algorithms for such quick analysis are called Level 1 (L1) triggers and are first part of trigger chain. After that selected events are passed to second part called High Level Triggers (HLT). They have more time for event selection thus tend to be more complicated algorithms.

In this analysis the interest is on events with forward proton in either side A or C. This can be done on L1 with e.g. L1AFP$\_$A$\_$OR$\_$C~\footnote{Trigger logic: [A Far and A Near] or [C Far and C Near]} trigger. EXC JJ events characterize also with jets in central region. They are more complicated objects, however they can be identified ether on L1 or HLT with e.g HLT$\_$j20~\footnote{Trigger logic: min. one jet with p$_T>$ 20 GeV} or L1$\_$J12~\footnote{Trigger logic: min. one jet with p$_T>$ 12 GeV}. 

Various combinations of L1 and HLT triggers were available in studied data, thus dedicated analysis was done in order to determine trigger providing best statistics after basic cuts for diffractive events\footnote{Those included: requirement of one vertex, two jets and proton either on side A or C and there were no problems during data taking (AFP Good Run List).}. It was found that trigger HLT$\_$j20$\_$L1AFP$\_$A$\_$OR$\_$C$\_$J12 is the best choice for this studies, which is combination of discussed previously triggers.

\section{Signal Selection}\label{sec:sig_selection}

Feasibility studies performed for the double-tagged events \cite{Trzebi_ski_2016} indicate that, due to the AFP acceptance, the jets should have the transverse momentum greater than 150 GeV. As the cross-section is steeply falling with jet $p_T$, such a measurement would require the data sample corresponding to the integrated luminosity of the order of inverse femtobarns. Events having jets of lower $p_T$ can be registered if the requirement of the forward proton tag is relaxed to only a single side tag (with the other proton presumable having too small $\xi$ to be detected in the AFP). Feasibility studies carried out for such semi-exclusive jet signature are described in \cite{Trzebi_ski_2015}. The selection proposed in this paper uses not only standard diffractive cuts (on Figure~\ref{fig:EXCJJ_XiRatio} marked as ``Standard") but also the kinematic relationship between the forward proton and the central jet systems as well as a veto concerning the activity outside the jet system. 

The kinematic relationship between forward and central systems are described using relative energy loss difference:
$$\xi_{rel} = (\xi_{p} - \xi_{dijet})/(\xi_{p} + \xi_{dijet}),$$
where $\xi_{p}/\xi_{dijet}$ are relative energy loss of proton and dijet\footnote{$\xi_{p}=1-\frac{E_{proton}}{E_{beam}}$ and $\xi_{dijet}^{\pm}=\frac{M_{dijet}}{2E_{beam}}\exp(\pm\eta_{dijet})$.}. Cut on this variable is marked on Figure~\ref{fig:EXCJJ_XiRatio} as ``$\xi_{ratio}$". The low activity outside dijet system is satisfied by the cuts on the number of tracks in pseudorapidity regions from edges of the detector to the edges of dijet (on Figure~\ref{fig:EXCJJ_XiRatio} marked as ``NTrkOutsideJet"), the number of tracks perpendicular to the leading jet in the azimuthal angle~\footnote{Track is considered perpendicular to the leading jet if $\frac{\pi}{3}<\Delta\phi<\frac{2\pi}{3}$ or $\frac{4\pi}{3}<\Delta\phi<\frac{5\pi}{3}$, where $\Delta\phi$ is the difference between the azimuthal angle of the track and the leading jet.} (on Figure~\ref{fig:EXCJJ_XiRatio} marked as ``NTrkPerpendicularJet") and the multiplicity of reconstructed clusters cells produced in the pseudorapidity range of $2.5 < |\eta| < 4.9$ (on Figure~\ref{fig:EXCJJ_XiRatio} marked as ``10>FCl on proton side" and ``max 1 Fcl''). Please note that the cut on forward clusters could not be implemented in run 336505 due to problems during data taking.
    


It is worth checking the behaviour of the $\xi_{rel}$ when a~selection is consecutively applied. Corresponding distributions for runs 336505 ($\mu \sim 0.04$) and 331020 ($\mu \sim 1$) are shown in Fig. \ref{fig:EXCJJ_XiRatio}. One could assume that the application of the selection criteria will make the shape of this distribution to peak around 0, as expected from \cite{Trzebi_ski_2015}. In the case of runs taken with $\mu\geq 1$ evidently, this is not the case. In the run taken with the lowest pile-up, the enhancement is visible already from the beginning.

\begin{figure}[!htbp]
    \centering
    \includegraphics[width=0.49\textwidth]{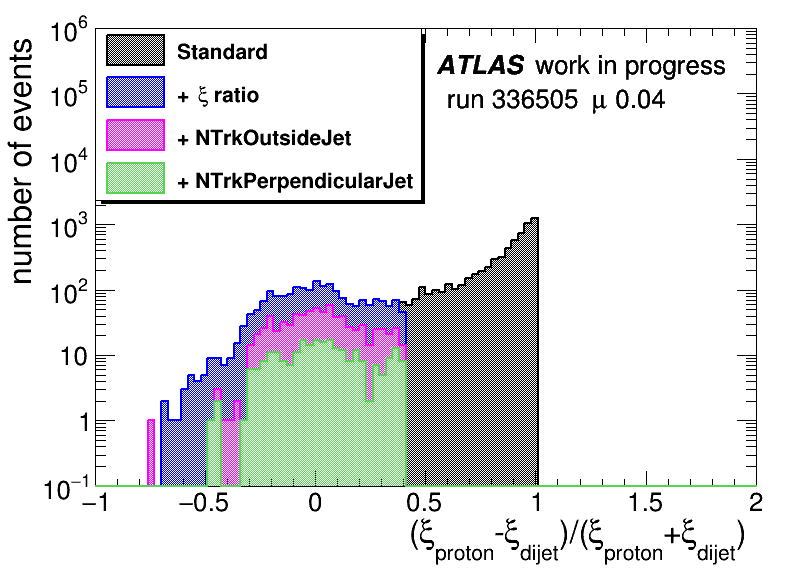}
    \includegraphics[width=0.49\textwidth]{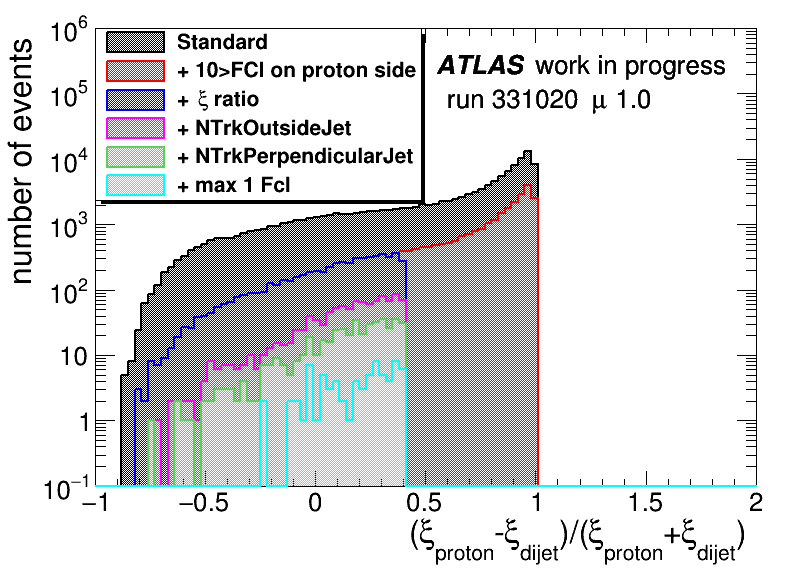}
    \caption{The distribution of the relative energy loss difference, $\xi_{rel} = (\xi_{p} - \xi_{dijet})/(\xi_{p} + \xi_{dijet})$ for runs 336505 ($\mu \sim 0.04$; \textbf{left}) and 331020 ($\mu \sim 1$; \textbf{right}). Various colours represent situations after a given, consecutive cut.}
    \label{fig:EXCJJ_XiRatio}
\end{figure}

From Figure~\ref{fig:EXCJJ_XiRatio} one could see that around 100 (in fact exactly 11 with proton on side A and 90 with proton on side C) events in~run 336505 passed the selection. From them, two are taken and shown as ``event displays'' in Figure~\ref{fig:EXC_JJ_event_displays} where the $\eta-\phi$ map of the ATLAS detector containing positions of the reconstructed objects is shown. The position of the reconstructed jets is shown as orange circles, with the value of the transverse momentum displayed next to it. The blue squares represent the clusters while the green triangles the tracks. The circle radius is proportional to the energy of the cluster or the transverse momentum of a jet/track.

The left plot shows a ``very clean'' event where most of the $\eta-\phi$ map is deprived of good-quality clusters and tracks. The right plot present case where three jets are present. Here two jets were produced very close to each other and their combined transverse momentum is of similar value to the third jet. It is still possible that those smaller jets should be in fact one but the jet reconstruction algorithm split it into two.

\begin{figure}[!htbp]
    \centering
    \includegraphics[width=0.49\textwidth]{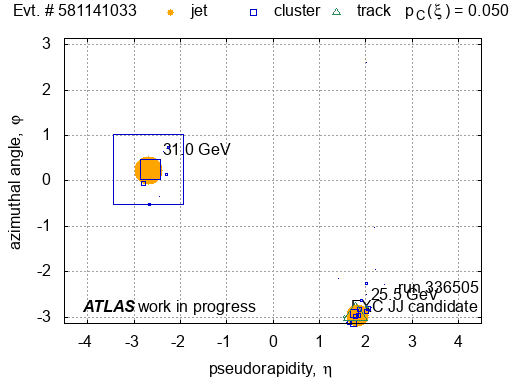}
    \includegraphics[width=0.49\textwidth]{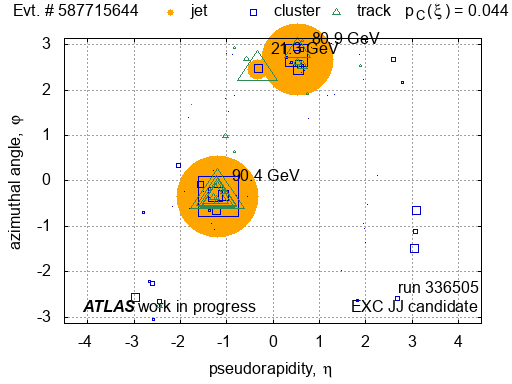}
    \caption{Displays of two events from run 336505 ($\mu \sim 0.04$) fulfilling the selection criteria for the exclusive jets.}
    \label{fig:EXC_JJ_event_displays}
\end{figure}

\section{Summary}

Following the suggestions from the feasibility studies~\cite{Trzebi_ski_2015}, the kinematic connection between the proton and jet system was used. The distribution of the relative energy loss difference, $\xi_{rel}$ , was expected to be peaked at 0 for the signal events. Quite surprisingly, this was the case only for data taken with the lowest pile-up, indicating that the background contribution in the case of other samples was much higher than expected. Finally, the most interesting cases, passing the exclusivity cuts, from the mentioned data set were selected and shown as event displays. 


\ResCnt

\maketitle

\begin{abstract}

The ATLAS Forward Proton Detector (AFP) at the LHC identifies events in which one or two protons emerge intact from the proton-proton collisions.  For this analysis, we looked at the single-diffractive production of charmed mesons.  In this case, only one of the protons interact and breaks up, and the second proton remains intact.  We analysed data from the AFP and plotted the distribution of mass difference for reconstructed $D^*(\pm)$ mesons produced in pp collisions at $\sqrt s$ = 13.6 TeV. We looked at the decay of a positively charged $D^*$ meson into a $D^0$ meson and slow, positively charged pion $(\pi^+_{slow})$, which then further decays into a negatively charged kaon $(K^-)$ and a positively charged pion $(\pi^+)$.  We did this to show the expected sensitivity for this kind of measurement using a small dataset and loose selection criteria.
\end{abstract}

\section{Introducing the AFP Detector}
The ATLAS Forward Proton (AFP) project tags and measures the position and angle of very forward protons.  From the position and angle at the AFP detector, we calculate the energy and $p_T$ of the protons. The detectors are placed about 200 m on either side of the interaction point in the central detector \cite{AFP_1}.  This detector is therefore useful in processes where one or two protons remain intact, better known as diffractive scattering.  The AFP detector consists of two trackers on either side of the interaction point.  Each tracker consists of 4 silicon planes providing precise position measurement by utilizing the 3D pixel technology.  The trackers are housed in movable Roman Pot devices that are used to approach the beam line as close as 2 - 4 mm.  The design features a ultra-high vacuum (UHV) enclosure, with thin floor and walls, thus limiting the material in the beam aperture \cite{AFP2}.

\section{Diffractive Events}
Diffractive events are characterised by at least one of the two incoming protons emerging from the interaction intact or excited into a low-mass state, with only a small energy loss.  We note that there are different classes of diffraction, including single diffractive dissociation (SDD),  central diffraction, double-diffractive dissociation and elastic collisions.  For this project, we only required a single proton in the AFP detector. \\
\\
Diffractive events can be explained by the exchange of a virtual object, the so-called Pomeron.  In the most naive approach within QCD, the Pomeron consists of 2 gluons, but it can have a structure more consistent with a so-called gluon ladder.\\
\\
In diffractive events, no hadrons are produced in a large rapidity range adjacent to the scattered proton, yielding a large rapidity gap (LRG) \cite{IFJ}.  The LRG can be used to identify diffractive events, but it is preferable to measure the scattered proton because of statistical fluctuations and signal overlap due to pile-up, as well as limited detector acceptance.

\section{Physics Motivation}
In charm quark production, the mass of the produced system is high enough that perturbative Quantum Chromodynamics (QCD) approximations are applicable.  Perturbative QCD is not applicable at lower masses, as the non-diminishing higher order terms inflate the computational complexity beyond achievable calculation.  As production cross section drops with increasing mass of the final state, the relatively low mass of the $D$ meson allows for high statistical significance of its experimental measurement.  As shown in this study, it is feasible to study the diffractive production of charmed mesons even in the low-luminosity dataset, allowing for a significant improvement of signal-to-background ratio.

\section{Analysis Results}
$D$ mesons are the lightest mesons containing charmed quarks or anti-quarks. For this project, the analysis was performed on $D$ mesons that decay into kaons and pions.  While the associated branching ratios are only on the level of a few \%, this decay channel offers the cleanest experimental signature.\\

We followed a similar analysis procedure as set out in references \cite{IFJ} and \cite{QCD}.  We looked at events with one intact tagged proton (a forward proton), and we look for a $D^*$ meson in those events.  The $D^*$ meson needs to decays into a $D^0$ meson and a soft pion.  The $D^0$ meson then decays into a kaon and pion of opposite charge.  

\[D^*(\pm) \to D^0\pi^{\pm}_{slow} \to (K^\mp \pi^\pm)\pi^\pm_{slow}\]

This channel is called the "golden channel" \cite{GC} because of the reduction in background and ease of identification of the $D$ meson candidates.  We only looked at charged particle tracks and we only used charge to differentiate between them.  This leads to a large combinatorial background because we take all possible combinations, even though there are other processes happening and there are other particle species that are being considered.  The candidates with the slow pions provide the most unique signature and significantly reduced the combinatorial background, making up for the relative rarity of the process occurring. \\
\\
We reconstructed the mass difference between the $D^*(\pm)$ mesons and the $D^0$ mesons in Fig. 1.  In each event, a pair of tracks from oppositely charged particles, each with $p_T > 1$ GeV, were combined to form the $D^0$ candidates.  An additional track with lower momentum ($p_T>0.25$) GeV was then combined with the $D^0$ candidate to form the $D^*$ meson.  We calculated the invariant mass of the $D^0$ mesons by assuming the mass of the kaons and pions for each track.  The additional low momentum track was assigned the pion mass.  The pions had to have the opposite charge of the kaon.  The $D^*$ mesons are identified by studying the distribution of the $D^\pm$ and $D^0$ invariant mass difference:

\[\Delta m = m(K\pi\pi_{slow})-m(K\pi)\]

\begin{figure}
\centering
\includegraphics[width=0.85\textwidth]{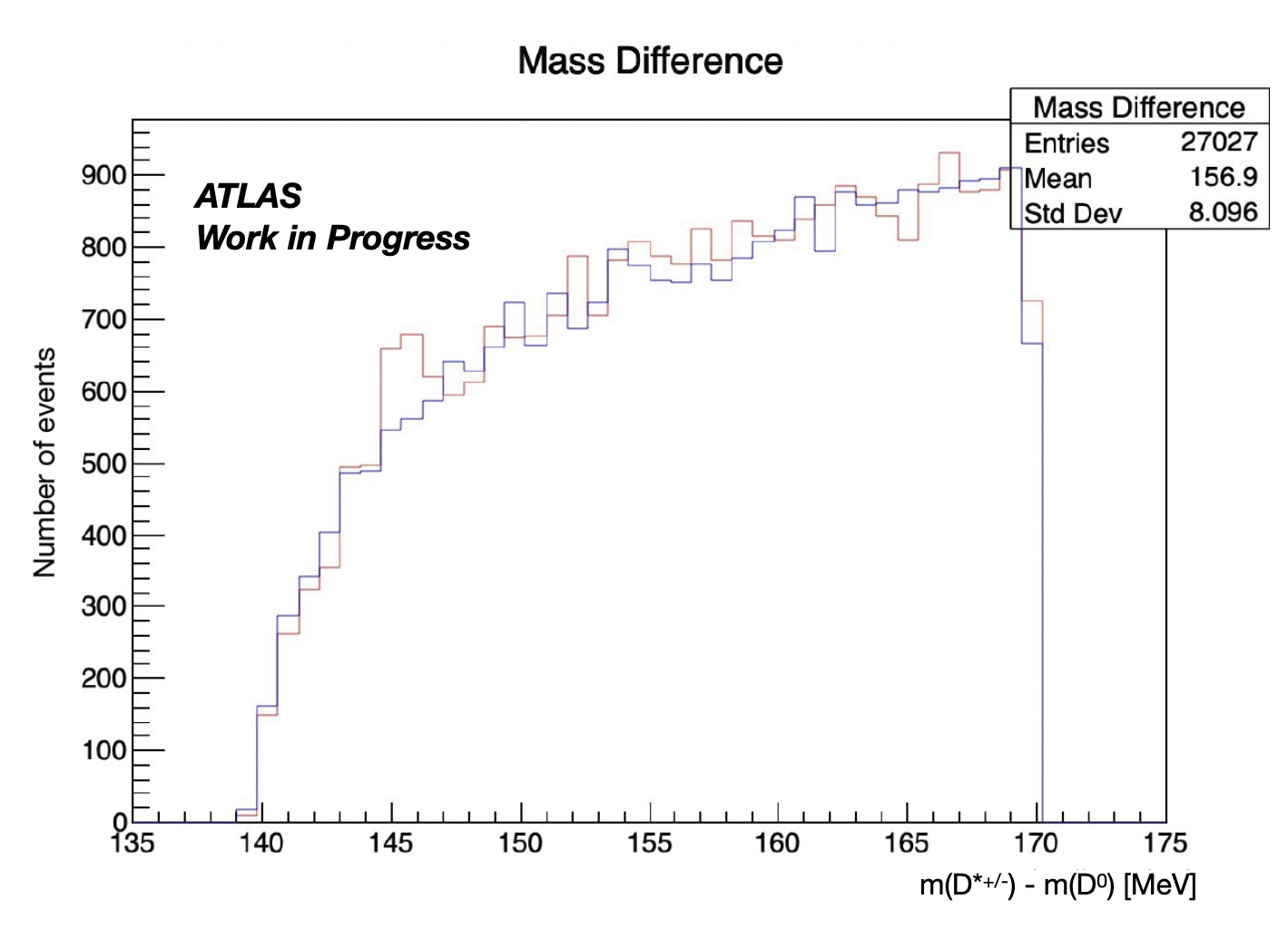}
\caption{\label{fig:plot}Mass difference of the reconstructed $D$ mesons.}
\end{figure}

In Fig. 1, the blue histogram is plotted with the incorrect charge combination.  The incorrect charge combination probes the combinatorial background and the excess at $\sim$ 145 MeV shows the $D^*$ signal.   If there were no $D$ mesons present, the distribution for the correct charge combination and the incorrect charge combination would be the same, and no excess would be observed.  The excess can therefore be interpreted as a signal and under this assumption we can calculate the cross-section.  \\
\\
Note that when selecting events, we were not using the topology of the decay process to constrain events (vertexing).  As a next step in the project, we want to check to see that the two tracks of the $D^0$ candidate intersect at a single vertex, and that this $D^0$ trajectory intersects with the third track to produce the $D^*$ candidate.  

\section{Summary}
This pilot analysis looked at the single-diffractive production of charmed mesons at the AFP detector.  The mass difference of the $D^*$ candidate and the $D^0$ candidate for two charge combinations was plotted, and a clear peak was seen for the correct charge combination.

\ResCnt

\maketitle
\vspace*{-1.2cm}
\begin{center}
{\small on behalf of the LHCb collaboration \cite{LHCb:2022mzw}}
\end{center}

\begin{abstract}
The first observation of the \LbDsP decay is presented using proton-proton collision data collected by the \lhcb experiment at the centre-of-mass energy of ${\sqs=13}$ \tev, corresponding to a total integrated luminosity of $6\invfb$. Using the $\LbLcPi$ decay as the normalisation mode, the branching fraction of the \mbox{$\LbDsP$} decay is measured to be ${\BF(\LbDsP)=}$ ${\BFLbDspPaper}$, where the first uncertainty is statistical, the second systematic and the third due to uncertainties corresponding to the branching fractions of the $\LbLcPi$, $\DsKKPi$ and $\LcPKPi$ decays.
\end{abstract}

\section{Introduction}
In the Standard Model (SM) of particle physics, the Cabibbo-Kobayashi-Maskawa (CKM) mechanism describes how the weak interaction eigenstates are related to the mass eigenstates of the quarks and determines the interaction strengths among quarks via the weak interaction~\cite{Cabibbo:1963yz,Kobayashi:1973fv}.The CKM-matrix element describing the $\bquark\to\uquark$ transition, $\Vub$, is the element with the smallest and most poorly determined magnitude. Better knowledge on $|\Vub|$ provides a valuable contribution for testing to check the consistency of the SM~\cite{UTfit-UT}.

The $\LbDsP$ decay\footnote{Inclusion of charge-conjugated modes is implied unless explicitly stated.} is a weak hadronic decay that proceeds through a $\bquark \to \uquark$ transition. A single leading-order diagram contributes to this process, shown in Fig.~\ref{fig:feyn-lb2dsp}. The $\LbDsP$ branching fraction is proportional to $|\Vub|^2$. 

\begin{figure}[htb!]
    \centering
    \vspace{+0.1cm}
    \includegraphics[width=0.7\textwidth]{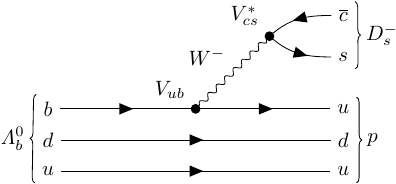}
    \caption{
    Tree diagram contributing to the \LbDsP decay.
    }
    \label{fig:feyn-lb2dsp}
\end{figure}
Study of this decay provides a measure to address the calculations of the branching fraction of the $\BdDsPi$ decay~\cite{LHCb-PAPER-2020-021}, which proceeds with the same tree-level transition as the $\LbDsP$ decay, leading to a similar expression for the branching fraction, except for the form factor and nonfactorisable effects. 

The paper~\cite{LHCb:2022mzw} presents the first observation and branching fraction measurement of the \mbox{$\LbDsP$} decay using proton-proton ($pp$) collision data collected with the LHCb detector at the centre-of-mass energy of $13\tev$ and corresponding to an integrated luminosity of $6\invfb$. The data taken in Run~2 of the Large Hadron Collider (LHC) between 2015 and 2018 are used. The $\LbLcPi$ decay is used as a normalisation channel because it is topologically similar to the signal decay and has a relatively high branching fraction. Candidates of $\LbDsP$ ($\LbLcPi$) decays are reconstructed using the final-state particles of the $\DsKKPi$ ($\LcPKPi$) decay. The branching fraction of $\LbDsP$ is determined using the following equation:
\small
\begin{equation}
    \BF(\LbDsP) = \BF(\LbLcPi) \dfrac{N_{\LbDsP}}{N_{\LbLcPi}} \dfrac{\epsilon_{\LbLcPi}}{\epsilon_{\LbDsP}} \dfrac{\BF(\LcPKPi)}{\BF(\DsKKPi)}\ ,
    \label{eq:BF}
\end{equation}
\normalsize
where $N_{X}$ is the measured yield of decay $X$ and $\epsilon_{X}$ is the efficiency of the candidate reconstruction and selection.


\section{Detector and simulation}
\label{sec:detector}
The \lhcb detector~\cite{LHCb-DP-2008-001} is a single-arm forward spectrometer covering the \mbox{pseudorapidity} range $2<\eta <5$, designed for the study of particles containing \bquark or \cquark quarks, e.g.\ via accessing the particle identification (PID) information with help of two ring-imaging Cherenkov detectors. Simulations are required to calculate reconstruction and selection efficiencies, and to determine shapes of invariant-mass distributions. Toolkits such as \pythia with a specific \lhcb configuration~\cite{LHCb-PROC-2010-056}, \evtgen and \geant, as described in Ref.~\cite{LHCb-PROC-2011-006}, are used.

\section{The invariant-mass fits}
\label{sec:selection-mass-fits}

The $\LbDsP$ (\LbLcPi) decay is reconstructed by selecting $\DsKKPi$ (\mbox{$\LcPKPi$}) candidates and combining them with a proton (charged pion), which is referred to as the companion particle. Candidates that have been selected by the trigger requirements are subject to further offline selection to reduce the background contributions, as described in Ref.~\cite{LHCb:2022mzw}.

The yields of the signal $\LbDsP$ and normalisation $\LbLcPi$ channels are determined using unbinned maximum-likelihood fits to the $\Dsm\proton$ and $\LcPi$ invariant-mass distributions, respectively.
The candidate samples from different years of data-taking and magnet polarities are combined in the fits. Parametrisations of the signal components are obtained from fits to samples of simulated candidates. The residual combinatorial background contribution is modelled using analytic functions. The background shapes are determined from simulation or described analytically~\cite{LHCb:2022mzw}.

The fit to the invariant-mass distribution of the $\LbDsP$ candidates is shown in Fig.~\ref{fig:lb2dsp}. A clear $\LbDsP$ signal peak is visible, corresponding to a yield of $\YieldLbDsp$ events, where the uncertainty is statistical. This result constitutes the first observation of this decay.

\begin{figure}[htb!]
   \begin{center}
        \includegraphics[width=0.99\textwidth]{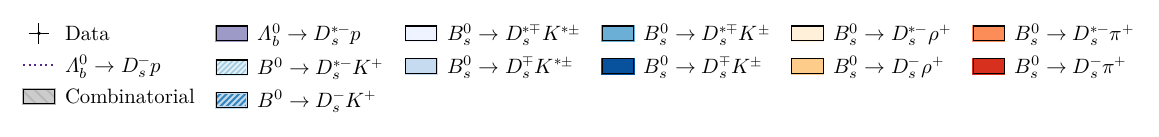}\\
        \includegraphics[width=0.74\textwidth]{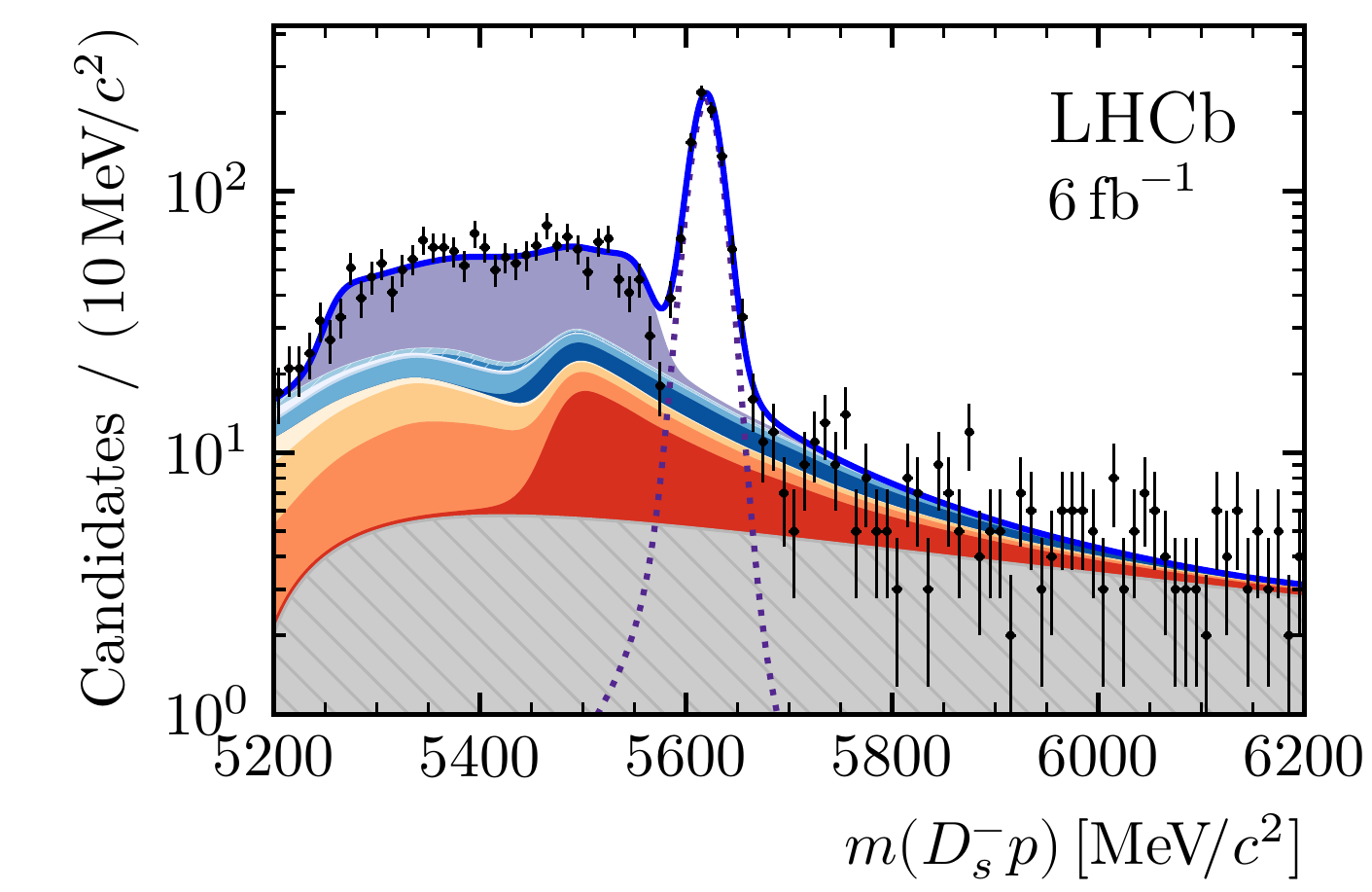}
    \end{center}
    \vspace{-0.5cm}
	\caption{Invariant-mass distribution of the $\LbDsP$ candidates, in  logarithmic scale, where the fit projections of the signal and background contributions are overlaid. The individual components in the fit are illustrated in the legend.}
	\label{fig:lb2dsp}
\end{figure}

\section{Results and conclusions}
\label{sec:results}

The branching fraction of $\LbDsP$ is determined using the selection efficiencies and the obtained yields of the $\LbDsP$ and $\LbLcPi$ decays, as in Ref.~\cite{LHCb:2022mzw}.
An external input for the $\LbLcPi$, $\LcPKPi$ and $\DsKKPi$ branching fractions is given in Table~\ref{tab:results}.

\begin{table}[htb!]
    \small
    \centering
    \caption{The obtained signal yields and efficiencies of the $\LbDsP$ and $\LbLcPi$ decays, as well as the branching fractions used for this measurement~\cite{PDG2022}. The uncertainty on the signal yields and efficiencies is statistical.}
    \label{tab:results}
    \begin{tabular}{lcc}
    \toprule
                    & $\LbDsP$                & $\LbLcPi$         \\
    \midrule
    Yield           & $\YieldLbDsp $          & $\YieldLbLcPiPaper$    \\
    Efficiency      & $(\EffLbDspPerc)\%$      & $(\EffLbLcPiPerc )\%$   \\
    \midrule       
    $\BF(\LbLcPi)$  & \multicolumn{2}{c}{ $(4.9\phantom{0} \pm 0.4\phantom{0})\times 10^{-3}$ ~\cite{PDG2022}} \\
    $\BF(\DsKKPi)$  & \multicolumn{2}{c}{ $( 5.38 \pm 0.10) \times 10^{-2}$ ~\cite{PDG2022}} \\
    $\BF(\LcPKPi)$  & \multicolumn{2}{c}{ $( 6.28 \pm 0.32 ) \times 10^{-2}$ ~\cite{PDG2022}} \\
    \bottomrule
    \end{tabular}
\end{table}  
The branching-fraction ratio of the $\LbDsP$ and $\LbLcPi$ decays is found to be
\begin{equation*}
    \frac{\BF(\LbDsP)}{\BF(\LbLcPi)} = \RatioBFPAPER \ ,
    \label{BF_ratio}
\end{equation*}
where the first uncertainty is statistical, the second systematic and the third due to the uncertainty on the $\DsKKPi$ and $\LcPKPi$ branching fractions. 
The branching fraction of $\LbDsP$ decay is obtained to be:
\begin{equation*}
    \BF(\LbDsP) = \BFLbDspPaper \ ,
    \label{eq:BF_Lb2Dsp}
\end{equation*}
where the third uncertainty is due to the uncertainty on the $\LbLcPi$, $\DsKKPi$ and $\LcPKPi$ branching fractions. This measurement is limited by the uncertainty on the $\LbLcPi$ branching fraction, which is dominated by the precision on the ratio of hadronisation fractions $f_{\Lb}/f_{d}$.

In summary, the first observation of the $\LbDsP$ decay and its branching fraction measurement are reported. Additionally, the branching fraction ratio of the $\LbDsP$ and $\LbLcPi$ decays is determined. This measurement will serve as input for future studies of factorisation in hadronic $\Lb$ decays.

%
%
%
%
%
%
%
%
%
%
%
%
%

\end{papers}

\end{document}